\documentclass[%
 reprint,
 amsmath,amssymb,
 aps,
prb,longbibliography
]{revtex4-2}

\usepackage{graphicx}
\usepackage{dcolumn}
\usepackage{bm}
\usepackage{braket}
\usepackage{amsmath}
\usepackage{amssymb}
\usepackage{feynmp}
\usepackage{graphics}
\usepackage{color}
\usepackage[hidelinks]{hyperref}
\hypersetup{colorlinks=false}

\usepackage[dvipsnames]{xcolor}
\usepackage[normalem]{ulem} 

\DeclareMathOperator{\Tr}{tr}
\renewcommand{\b}[1]{\mathbf{#1}} 

\begin{document}


\title{Itinerant ferromagnetism in transition metal dichalcogenides moiré superlattices}
\author{Pawel Potasz$^{1}$}
\author{Nicol\'as Morales-Dur\'an$^{2}$} 
\author{Nai Chao Hu$^{2}$}
\author{Allan H. MacDonald$^{2}$}

\affiliation{$^1$Institute of Physics, Faculty of Physics, Astronomy and Informatics, Nicolaus Copernicus University, Grudziadzka 5, 87-100 Toru\'n, Poland}
\affiliation{$^2$Department of Physics, University of Texas at Austin, Austin, Texas, 78712, USA}

\date{\today}

\begin{abstract}
Moir\'e materials are artificial crystals formed at 
van der Waals heterojunctions that have emerged as a highly tunable platform
that is able to realize much of the rich quantum physics of electrons in atomic scale solids,
and in several cases even new quantum phases of matter. 
Here we use finite-size exact diagonalization methods to explore the physics of 
single-band itinerant electron ferromagnetism in semiconductor moiré materials.  We predict 
where ferromagnetism is likely to occur in triangular-lattice moiré systems, 
and where it is likely to yield the highest Curie temperatures.
\end{abstract}

\maketitle










\section{Introduction}
Moir\'e materials have already been established as hosts of 
Mott \cite{HubbardCornell,ExperimentMITCornell,ExperimentMITColumbia} and topological insulators \cite{QAHCornell},
a rich variety of magnetic states \cite{FrustratedWignerMott, Xu_Ferromagnetism, ciorciaro2023kinetic, tao2023observation},
and recently even fractional Chern insulators \cite{cai2023signatures,zeng2023thermodynamic}.
 They also provide an alternative platform for studies of itinerant electron ferromagnetism \cite{tao2023observation, JiaBowen_itinerFerro_Moire, Polaron_moire, FTLM_Vafek}.
Ferromagnets are many-electron ground states that break time-reversal but not translational symmetry, 
have finite macroscopic magnetization, and are more common in metals than in insulators.
Ferromagnetic metals exhibit a rich variety of interesting hysteretic magneto-resistive effects
that lie at the heart of spintronics \cite{bader2010spintronics} and are valuable for technology. 
Theoretical studies of metallic ferromagnetism in the context of simple one-band Hubbard models \cite{TasakiReview98,Hanisch_HubbFerro,Vollhardt98,MetallicFerroKeller,MetallicFerroKoga, lebrat2023observation, prichard2023directly,HTkineticDemler}, although rarely physically realistic,
have nevertheless helped provide an understanding of the necessary conditions to stabilize such ground states 
in crystalline materials.  The moiré material case, in which isolated bands are common,
offers the opportunity to compare theories of single-band itinerant electron ferromagnetism
directly with experiment.  

In this paper we use exact 
diagonalization methods (ED) to explore metallic ferromagnetism in the single-band triangular lattice moiré materials
realized in transition-metal dichalcogenide (TMD) heterobilayers \cite{wu2018hubbard,FengchengTMD1,Review1,Review2} such as WSe$_2$/MoSe$_2$ and WSe$_2$/WS$_2$. We 
predict where ferromagnetism is most likely to occur and where ferromagnetic
transition temperatures are maximized. The restriction of our study to the case in which a single band is
partially occupied and well separated from other bands \cite{NicolasWigner} is motivated by a technical 
consideration, namely the need to restrict the dimensions of the many-electron 
Hilbert spaces studied to manageable sizes \footnote{Hilbert space dimensions grow exponentially 
with the number of retained bands.}.
Metallic ferromagnetism is interesting in both single-band and multi-band systems. In the multi-band case  
local moments from one subset of bands that supply local Hunds magnetism can combine with  
large spin-stiffnesses supplied by another set of band that validate simple 
mean-field descriptions - using density functional theory for true atomic scale materials.
In contrast, single-band systems are often more difficult to understand, requiring non-perturbative approaches as the one we take here. Although it seems likely that the highest ferromagnetic 
transition temperatures that can be realized in moiré systems are in multi-band systems \footnote{
The highest critical temperature metallic ferromagnets in atomic scale crystals are after all
multi-band systems.} we nevertheless anticipate that 
scientific progress can be achieved by comparisons between theory and experiment 
across a broad range of band filling factors and band widths in the single isolated-band regime.

Our paper is organized as follows.  In Section ~\ref{sec:model} we specify the model that we study
- a triangular lattice moiré material model with the Hilbert space truncated to 
the lowest energy moiré band and interaction matrix elements calculated exactly. In Section ~\ref{sec:results}
we present our numerical results.  We examine three different ferromagnetism indicators that are 
available from finite-size ED calculations; i) ground state spin quantum numbers,
ii) magnon energy estimates from the total-momentum dependence of the low-energy many-body excitation spectrum
and iii) Lanczos spin-susceptibility calculations.
All are consistent with the notion that ferromagnetism occurs when the band filling factor 
of the lowest energy hole miniband is around $\nu \sim 3/4$.  We estimate that
Curie temperatures that can reach $T\sim 10$ K.  
Finally in Section ~\ref{sec:discussion} we summarize and discuss our findings, estimating conditions for which 
the single-band model is realistic. We conclude that the 
single-band approximation is not applicable at $\nu \sim 3/4$ 
in the TMD moir\'e materials studied experimentally to date, but that 
it can be realized by choosing systems with the strongest possible moir\'e potentials
and maximizing background screening of the Coulomb interaction. 

\section{Finite Size Moir\'e Material Model}
\label{sec:model}

In this paper we will focus on transition metal dichalcogenide
heterobilayer moiré materials \cite{wu2018hubbard} in which the topmost valence miniband is 
energetically isolated, so that holes only populate this band upon doping.
Because we are interested mainly in understanding where ferromagnetism has a substantial ordering temperature, 
we focus on the range of twist angles for which the topmost band is relatively dispersive. 
The single-particle part of the continuum model Hamiltonian describing these systems is \cite{wu2018hubbard}
\begin{align}
    \label{Continuummodel}
    H_0&=-\frac{\hbar^2}{2m^*}{\bm k}^2 +\Delta({\bm r}),\\
    \label{moirépotential}
    \Delta({\bm r})=&2V_m\sum_{j=1,3,5}\cos({\bm b}_j\cdot{\bm r}+\psi),
\end{align}
where the ${\bm b}_j$ are members of the first shell of moir\'e reciprocal lattice vectors and 
$m^*$, $V_{{\rm m}}$ and $\psi$ are heterojunction specific parameters.
The specific calculations we report on below take effective mass $m^*=0.35 \,m_0$, where $m_0$ is 
the rest mass of the electron, moir\'e modulation strength $V_{{\rm m}} = 25$ meV,
and moir\'e potential shape parameter \cite{FengchengTMD1} $\psi=-94^{\circ}$. These numerical values correspond to WSe$_2$/MoSe$_2$ heterobilayer moir\'es \cite{wu2018hubbard}. It is known \cite{DeepMoire_Columbia,wang2023fractional} 
that strain relaxation of the moir\'e pattern strengthens the moir\'e modulation potential,
an effect that can be incorporated approximately simply by increasing the value of $V_{{\rm m}}$. For this reason we take a slightly larger value for the moiré modulation than the one reported for the unstrained bilayer \cite{wu2018hubbard}. (Approximate 
scaling relations relating our results to those at larger values of $V_{{\rm m}}$ are explained in the 
discussion section.) 
\begin{figure}[h!]
\centering
\includegraphics[width=\linewidth]{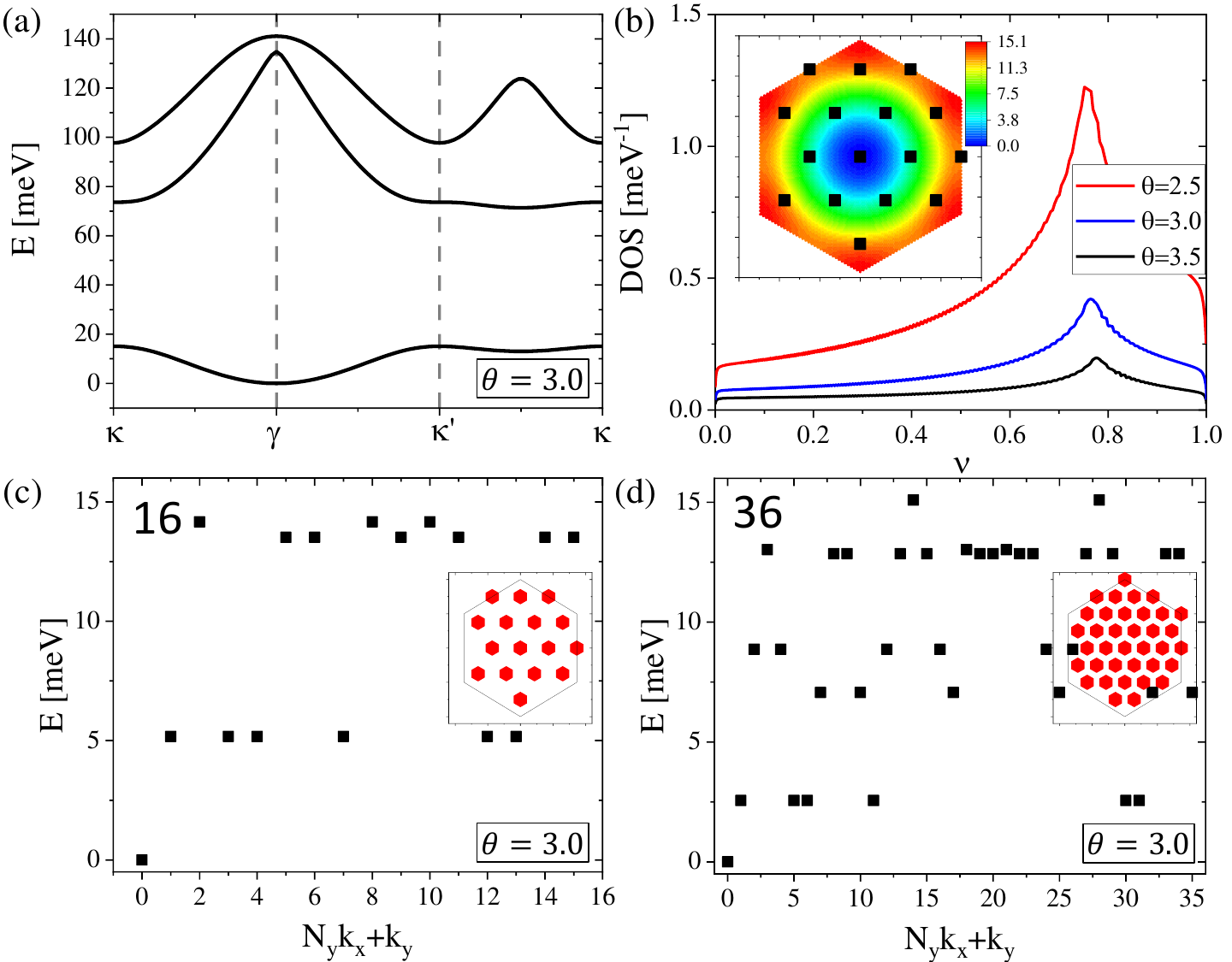}
\caption{(a) Particle-hole transformed (hole-picture) bandstructure of 
moiré TMD heterobilayers at twist angle $\theta = 3.0$,
moir\'e modulation strength $V_{{\rm m}} = 25$ meV and shape parameter $\psi=-94^{\circ}$.
Note that the lowest energy hole miniband is partially occupied
and isolated from the remote bands.
(b) Density of states (DOS) of the lowest energy hole miniband $\it{vs.}$ band filling $\nu$
for twist angles $\theta = 2.5$, $\theta = 3.0$, $\theta = 3.5$. The inset indicates the discrete momenta of a 
$M=16$ unit cell finite-size system within a color scale band contour plot for the $\theta = 3.0$ case. 
These bands have a van Hove singularity at energy $E_{\rm VH} \approx 15$ meV and band filling factor 
$\nu_{\rm VH} \approx 0.75$ in the thermodynamic limit, $M \to \infty$. 
(c) and (d): The discrete energies of the (c) $M=16$ and (d) $M=36$ finite size systems discussed in the 
text. }
\label{fig:Singlepart}
\end{figure}

Figures \ref{fig:Singlepart}(a) and (b) illustrate the implied moir\'e band structures
and densities-of-states.  The density-of-states maximum occurs at the energy of a 
saddle-point van Hove singularity (VHS) at band filling $\nu \approx 3/4$, where $\nu = \frac{N}{2M}$ with $N$ the number of valence band holes in the system (we call them particles from now on), and $M$ the number of moir\'e unit cells.
We will find that ferromagnetism occurs when the van 
Hove singularity is close to the Fermi level of the competing paramagnetic state.
The position of the van Hove singularity (VHS) shifts slightly to
larger band filling factors $\nu$ with increasing twist angle. 
The VHS is manifested in finite size calculations with $M$ unit cells  
by a bunching of discrete states in a small energy interval.  
In Fig. \ref{fig:Singlepart}(c) we show the discrete single-particle spectra of 
(c) $M=16$ and (d) $M=36$ meshes.  
When momentum space is discrete, the thermodynamic limit VHS results in a 
set of closely spaced discrete energies slightly below $E=15$ meV. 
When these states are occupied only by majority spins and all other states are 
doubly occupied the filling factor is $\nu = 0.72$ for $M=16$ and $\nu = 0.74$ for $M=36$ system sizes, respectively.
Note that single particle states at general momenta in the Brillouin-zone interior are 
six fold degenerate simply due to triangular lattice rotational symmetries; this property is responsible for 
the bunching near $E=5.0$ meV for $M=16$ and near $E=2.5$, $E=7.0$, $E=9.0$ meV for $M=36$.
($\gamma$ point (${\bf k} = 0$) states are non-degenerate and Brillouin zone corner states are doubly-degenerate - 
the degeneracy between $K$ and $K'$ points.) As is commonly recognized, the bunching of single particles 
energy levels has an impact on finite-size many-body results, and limits the types of conclusions that can be reached.
We will consider a variety of different finite size geometries, each with a corresponding 
discretization of the moiré Brillouin zone.
In order to correctly capture the VHS physics , we seek meshes that neither underrepresent nor overrepresent the 
associated high density of states close to $\nu = 0.75$.
In the Supplemental Materials (SM) \cite{Supplemental} we discuss how we choose finite-size geometries for the calculations discussed in the main text (see also Refs. \cite{TiltedM, AvellaSpringer, PhysRevB.90.245401} therein).

The full Hamiltonian is obtained by projecting the two-particle 
Coulomb interaction term to the topmost valence band shown in Fig. \ref{fig:Singlepart}(a):
\begin{align}
    \label{ManyBodyHamiltonian}
    H=  H_{\rm 0} + H_{\rm I}\nonumber\\ = \sum_{{\bm k},\sigma}\epsilon_{{\bm k}}c^{\dagger}_{{\bm k}\sigma}c_{{\bm k}\sigma}+\frac{1}{2}\sum_{\substack{i,j,k,l\\
    \sigma,\sigma^{\prime}}} \; V_{i,j,k,l}^{\sigma,\sigma^{\prime}} \; c^{\dagger}_{{\bm k}_i\sigma}c^{\dagger}_{{\bm k}_j\sigma^{\prime}} c_{{\bm k}_l\sigma^{\prime}}c_{{\bm k}_k\sigma},
\end{align}
where $c^{\dagger}_{{\bm k}\sigma}$ ($c_{{\bm k}\sigma}$) creates (annihilates) a particle with momentum ${\bm k}$ and spin $\sigma$, $\epsilon_{{\bm k}}$ are band energies, and the Coulomb matrix elements are given by 
\begin{align}
\label{InteractionME}
    V_{i,j,k,l}^{\sigma,\sigma^{\prime}}=\frac{1}{A}\sum^{'}_{\substack{{\bm G}_i,{\bm G}_j\\{\bm G}_k,{\bm G}_l}} \left(z^{*}_{{\bm k}_i,{\bm G}_i} z^{*}_{{\bm k}_j,{\bm G}_j}z_{{\bm k}_k,{\bm G}_k}z_{{\bm k}_l,{\bm G}_l}\right) \frac{2\pi e^2}{\epsilon\, q},
\end{align}
with $z_{{\bm k},{\bm G}}$  eigenstate coefficients obtained from diagonalization of Hamiltonian $H_{\rm 0}$  given by Eq. (\ref{fig:Singlepart}) in a basis of plane waves ${\bm G}$.
In Eq. (\ref{InteractionME}) $A$ is moir\'e unit cell area, momentum conservation implies that matrix elements are non-zero only
if ${\bm k}_i+{\bm k}_j = {\bm k}_k + {\bm k}_l$ modulo a moir\'e reciprocal lattice vector, the prime on the sum over the 
${\bf G}$'s implies that ${\bm k}_i+{\bf G}_i+{\bm k}_j+{\bm G}_j = {\bm k}_k + {\bm G}_k + {\bm k}_l+ {\bm G}_l$, and 
$q=|{\bm q}| = |{\bm k}_i+{\bf G}_i - {\bm k}_k - {\bm G}_k|$ is the momentum transfer.  As we have shown previously \cite{Nonlocal}, by working in a Wannier representation
the matrix elements can be reexpressed in terms of a single large parameter, the on-site Coulomb interaction
$U_{\rm 0}$, and a series of smaller parameters including non-local exchange, interaction assisted hopping, and 
longer range local interactions.  The strength of interactions depends on the value used for the 
effective dielectric constant $\epsilon$, which represents screening by the three-dimensional dielectric 
environment of the moir\'e system.  We return to this issue in the discussion section.  

\begin{figure}[h!]
\centering
\includegraphics[width=0.9\linewidth]{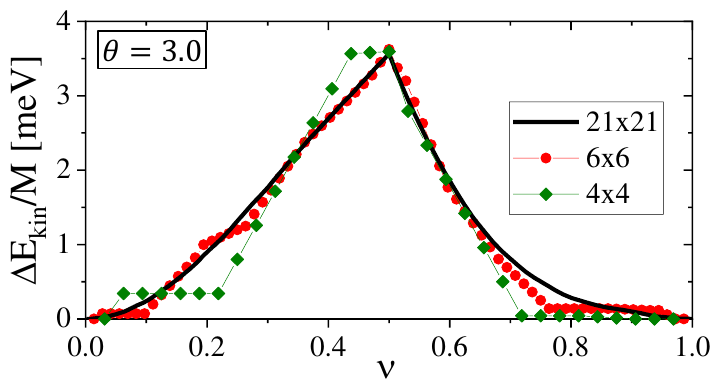}
\caption{Kinetic energy difference between maximal and minimal spin polarized states per moir\'e cell
$\Delta E_{{\rm kin}}/M = (E^{{\rm min}}_{{\rm kin}}(S^z_{\rm max}) - E^{{\rm min}}_{{\rm kin}}(S^z_{\rm min}))/M$
as a function of band filling factor $\nu=N/2M$ for $\theta = 3.0$.  }
\label{fig:Fig_kin}
\end{figure}

The physics of ferromagnetism is often viewed qualitatively as a competition between band energies,
which favor states with minimal spin-polarization and interaction energies, which favor spin-polarized 
states because many-electron wavefunctions must vanish when electrons with parallel spins approach each other, thereby 
avoiding strong repulsive interactions.  The gain in interaction energy per unit cell is often referred to as the 
Stoner energy $I$.  In Fig.~\ref{fig:Fig_kin} we compare finite size kinetic energies for 
single-Slater-determinant (SD) states with maximal and minimal spin-polarization in triangular lattice moir\'e materials,
$\Delta E_{{\rm kin}} = E^{{\rm min}}_{{\rm kin}}(S^z_{\rm max}) - E^{{\rm min}}_{{\rm kin}}(S^z_{\rm min})$, where the
superscripts `min' emphasize that the occupation numbers are chosen to minimize the kinetic energy subject to the spin-polarization constraint.
The energy difference per moir\'e period reaches its maximum when the band is half-filled because this is the 
filling factor with the maximum possible spin-polarization per moir\'e cell. The kinetic energy cost increases with twist angle $\theta$ because of increasing band widths, see SM \cite{Supplemental}.
Note that the kinetic energy cost of spin-alignment is, for the most part, reasonably well approximated 
at relatively small system sizes, and that the kinetic energy cost is very small for large band 
filling factors because of the VHS near the top of the first hole miniband. This is the filling factor regime where itinerant ferromagnetism might be expected.

\section{Exact Diagonalization results}
\label{sec:results}

We will discuss three different indicators for ferromagnetism that are
available from finite-size calculations.  First of all we consider the 
total spin quantum number of the finite-size many-electron ground state.   
The absence of spin-orbit coupling in our model allows a ferromagnet to be defined as a 
system in which the ground state total spin quantum number $S$ is extensive.  
We find that maximal spin-polarization is common in finite-size systems at band filling factors larger than
about $3/4$, and conclude that ferromagnetism will occur through much of this filling factor range.
In the following subsections we estimate the temperature to which 
ferromagnetism survives in two different ways: i) by extracting magnon-energies from the 
momentum dependence of the many-body excitation spectrum and ii) by extracting finite temperature 
Stoner energies $I$ from the temperature-dependent spin-susceptibilities calculated using 
finite-temperature Lanczos methods.

\begin{figure}[ht]
\centering
\includegraphics[width=\linewidth]{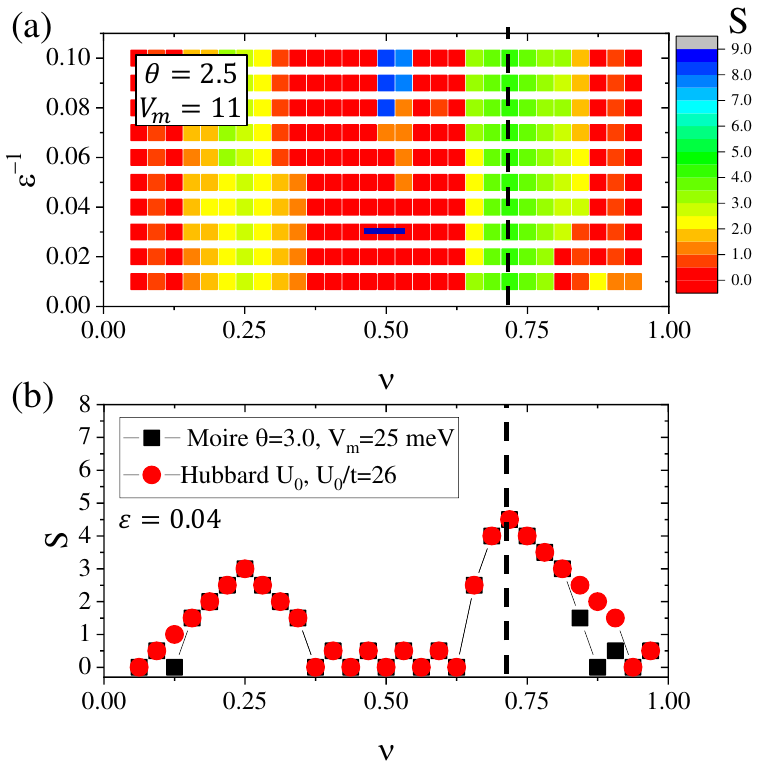}
\caption{The ground state total spin $S$ as a function of filling factor $\nu$ from exact diagonalization calculations for a system with $M=N_{\rm 1}\times N_{\rm 2} = 16$ unit cells ($N_{\rm 1}$ and $N_{\rm 2}$ are defined in SM \cite{Supplemental}). 
(a) Spin polarization map: Total spin as a function of filling factor $\nu$ and dielectric constant $\epsilon$ for twist angle $\theta = 2.5$ and moir\'e potential strength $V_{{\rm m}} = 11$ meV. A horizontal blue line labels the metal-insulator transition at half-filling \cite{MoireMIT}.
(b) Comparison of the ground state spin polarization of the moiré continuum Hamiltonian and the corresponding on-site Hubbard model 
for dielectric constant $\epsilon^{-1} = 0.04$, twist angle $\theta = 3.0$, moir\'e strength $V_{{\rm m}} = 25$ meV and 
moir\'e shape $\psi=-94^{\circ}$. A dashed line indicates the position of the van Hove singularity for finite size mesh. }
\label{fig:SpinvsFill}
\end{figure}

\subsection{Ground-state spin}

We first assess the tendency toward ferromagnetism by comparing ground state energies in different total 
spin $S$ sectors.  Typical results are summarized in Fig.~\ref{fig:SpinvsFill}(a), where we plot ground state 
spin quantum numbers vs $\nu$ and the interaction strength parameter $\epsilon^{-1}$.  Large ground state 
spins appear in several different regimes in this plot.  First of all they appear at small band filling factors 
and weak interactions.  We view ferromagnetism in this regime as an artifact of the symmetry-related 
momentum-space shell degeneracy of the finite-size mesh used to produce these results, which we have illustrated in 
Fig.~\ref{fig:Singlepart}.  Secondly, ferromagnetism is 
seen near half-filling of the band at large interaction strengths.   
The ground state at $\nu=1/2$ for this range of interaction parameters is an interaction-induced insulator (the blue line in Fig.~\ref{fig:SpinvsFill}(a) labels a metal-insulator transition estimated from the charge gap calculations using a definition: $\Delta_C = E_{\rm tot}(N+1) + E_{\rm tot}(N-1) - 2E_{\rm tot}(N)$, see also. Ref. 23), but the 
ground state is ferromagnetic rather than antiferromagnetic because spatially indirect exchange interactions
($\propto \epsilon^{-1}$) exceed antiferromagnetic superexchange interactions ($\propto \epsilon$).  
The property that Mott insulators are sometimes ferromagnetic in moir\'e materials has been discussed previously \cite{Nonlocal}. Our main interest here is in the very robust 
ferromagnetic states that appear near band filling $\nu=3/4$, where the ground state is metallic.
In Fig.~\ref{fig:SpinvsFill}(b) we plot the ground state spin {\it vs.} $\nu$ in the
moderate interaction strength regime, where non-local exchange is unimportant, 
demonstrating that its value is unchanged when the interaction model is truncated to include only the on-site 
Hubbard-like Coulomb interaction term. In SM \cite{Supplemental} we show that the magnetic 
competition in the insulating  state at $\nu=1/2$ is shifted in favor of antiferromagnetism with increasing twist angle, 
as expected since larger band widths imply stronger superexchange interactions.

\begin{figure}[h!]
\centering
\includegraphics[width=\linewidth]{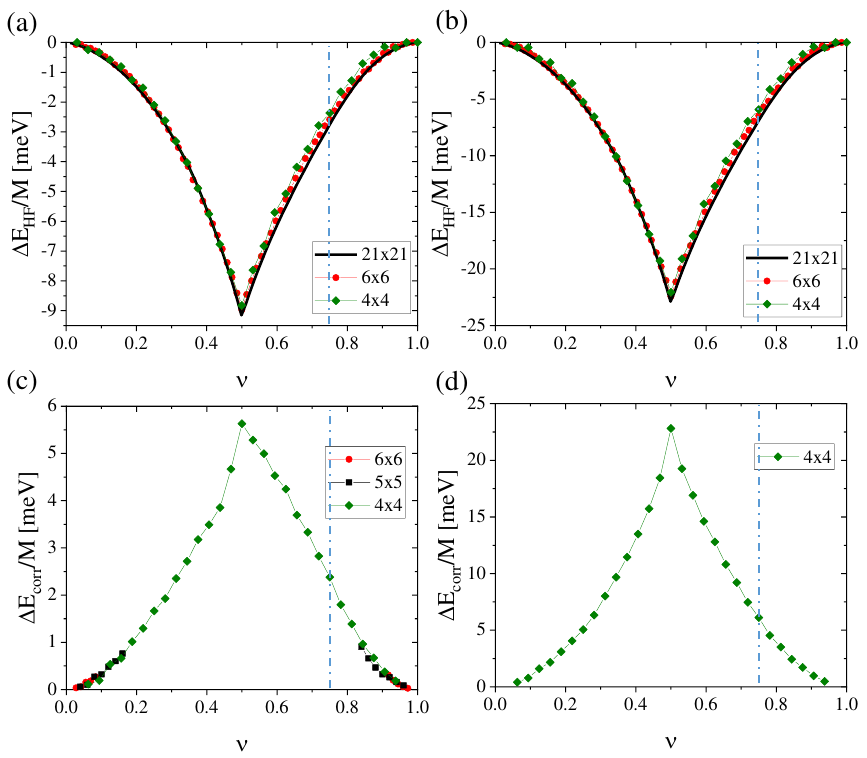}
\caption{Exchange energy and correlation energy difference between maximally $S_{\rm max}$ and minimally $S_{\rm min}$ spin polarized states normalized per moir\'e unit cell.  
These plots are based on finite-size ED calculations for $M=16$ and on
non-self-consistent Hartree-Fock, single Slater determinant SD, for $M=441$. A dashed line indicates the position of the van Hove singularity. (a) and (c) for interaction strength $\epsilon^{-1}=0.04$ and (b) and (d) for interaction strength $\epsilon^{-1}=0.1$. 
These plots are for twist angle $\theta = 3.0$, moir\'e modulation strength $V_{{\rm m}} = 25$ meV, and potential shape $\psi=-94^{\circ}$.}
\label{fig:ExchCorrel}
\end{figure}

In Fig.~\ref{fig:ExchCorrel} we analyze the competition between ferromagnetism and paramagnetism by 
partitioning the total energy into four different contributions: 
kinetic energy $E_{{\rm kin}}$, Hartree energy $E_{{\rm H}}$, Fock (exchange) energy $E_{{\rm exch}}$, 
and correlation energy $E_{{\rm corr}}$.  Convergence to the thermodynamic limit is easily obtained 
for the first three terms, whereas the fourth part, the correlation energy, must
be estimated from finite-size calculations and extrapolated to the thermodynamic limit.  
For the purposes of the qualitative point that we wish to make in this paragraph, we define the sum of the 
first three terms as the expectation value of the full Hamiltonian in the single Slater determinant (SD)
state constructed by occupying the lowest energy single-particle states for a given spin-polarization.
We define the mean-field interaction energy difference $\Delta E_{{\rm HF}}= \Delta E_{\rm H} + \Delta E_{\rm exch}$ between maximally and minimally spin polarized SD states by subtracting the kinetic energy contribution to the energy difference:
\begin{align}
    \label{exch}
    \Delta E_{{\rm HF}} =\Delta E_{{\rm SD}} - \Delta E_{{\rm kin}}.
\end{align}
Note that $\Delta E_{{\rm HF}}$ accounts for the fact that the shape of the charge distribution 
within the unit cell is different in the spin-polarized and unpolarized state, an effect that is absent in the 
Hubbard model. Because of this effect, the lowest energy SD state is not always the one constructed 
from the lowest energy single-particle states. In the SM \cite{Supplemental} we show results for
$\Delta E_{{\rm exch}}$ obtained from multi-band self-consistent Hartree-Fock calculations.
These energies have larger negative values because of the additional band-mixing degrees-of-freedom that 
are optimized. 

The correlation energy is defined as the 
difference between the ED ground state energy and the lowest energy SD ground state energy in a given spin sector with subtracted kinetic energies contributions
\begin{align}
    \label{correl}
    E_{{\rm corr}}(S_{\rm max}) =& E_{{\rm tot}}(S_{\rm max}) -  E^{{\rm kin}}_{{\rm tot}}(S_{\rm max}) \\\nonumber -& (E_{{\rm SD}}(S_{\rm max}) - E^{{\rm kin}}_{{\rm SD}}(S_{\rm max})).
\end{align}
Here we used the following  definitions 
\begin{align}
E_{\rm tot}(S) &= \langle\Psi_{\rm GS}(S)|H_{\rm 0} + H_{\rm I} |\Psi_{\rm GS}(S)\rangle, \nonumber \\
E_{\rm tot}^{\rm kin}(S) &= \langle\Psi_{\rm GS}(S)|H_{\rm 0}  |\Psi_{\rm GS}(S)\rangle,\nonumber \\
E_{\rm SD}(S) &= \langle\Phi_{\rm GS}(S)|H_{\rm 0} + H_{\rm I} |\Phi_{\rm GS}(S)\rangle, \nonumber\\
E_{\rm SD}^{\rm kin}(S) &= \langle\Phi_{\rm GS}(S)|H_{\rm 0}  |\Phi_{\rm GS}(S)\rangle, \nonumber
\end{align}
where $|\Psi_{\rm GS}(S)\rangle$ is ED ground-state wave function in a total spin sector $S$ and $|\Phi_{\rm GS}(S)\rangle = \prod_{{\bf k}\sigma} c^\dagger_{{\bf k}\sigma}|0\rangle$ is lowest energy SD state. 
The correlation energy difference is
\begin{align}
    \label{Dcorr}
    \Delta E_{{\rm corr}} = E_{{\rm corr}}(S_{\rm max}) -  E_{{\rm corr}}(S_{\rm min}).
\end{align}
With the above definitions, the total energy difference is
\begin{align}
    \label{Dcorr2}
    \Delta E_{{\rm tot}} = \Delta E_{{\rm SD}} + \Delta E_{{\rm corr}}.
\end{align}


In Fig.~\ref{fig:ExchCorrel} we see that mean-field interaction energies
$\Delta E_{{\rm HF}}$ strongly favor spin-polarized states, and that the degree to which interactions favor spin-polarized
states is strongly reduced when correlations are included.  For the parameters of this calculation,
increasing the strength of interactions actually does not substantially increase the degree to which 
interactions favor spin-polarization.
This is precisely the problem in estimating where ferromagnetism occurs;
once correlations are strong, electrons avoid each other well even if they have the same spin, and even in metallic states.
Ferromagnetism is most likely
when one subset of states has a high density-of-states so that it is easily polarized, and the remaining states are strongly
dispersive so that correlations are suppressed. Conditions favorable for ferromagnetism are regularly achieved in 
multi-band systems, like the paradigmatic late $3d$ transition metals.  
In single-band systems somewhat less favorable 
conditions can be achieved by having a sharp maximum in the density-of-states.  For two-dimensional (2D) materials, maxima always appear at 
saddle points in the band structure.  It follows that single-band ferromagnetism in 2D moiré
materials is most likely when the Fermi level of the paramagnetic state is close to a saddle point in the band structure.

\begin{figure}[h!]
\centering
\includegraphics[width=\linewidth]{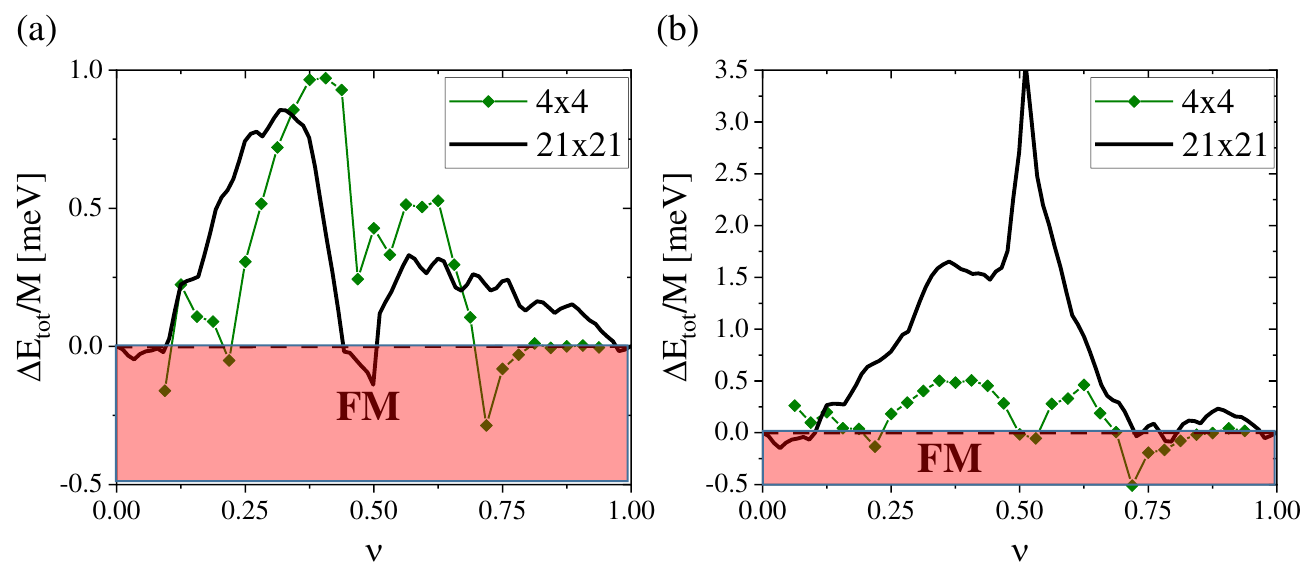}
\caption{The total energy difference between maximally $S_{\rm max}$ and minimally $S_{\rm min}$ spin polarized states $\Delta E_{{\rm tot}} = E_{{\rm tot}}(S_{\rm max}) - E_{{\rm tot}}(S_{\rm min}))$ per moir\'e unit cell for (a) $\epsilon^{-1}=0.04$ and for (b) $\epsilon^{-1}=0.1$. $\Delta E_{{\rm tot}}$ for $M=16$ is obtained from ED calculations and for $M=441$ from exchange energy and extrapolated correlation energy from ED. These results were obtained with model parameters
$\theta = 3.0$, $V_{{\rm m}} = 25$ meV and $\psi=-94^{\circ}$.} 
\label{fig:Etot}
\end{figure}

A typical result for the competition in total energy between fully spin polarized and depolarized states is 
summarized in Fig.~\ref{fig:Etot} where we see that ferromagnetism is most likely near $\nu=3/4$ as expected.
The Hartree-Fock theory results for the weaker of the two interaction strengths considered tell a cautionary tale 
about finite-size effects since they predict ferromagnetism for $M=16$ finite-size systems and paramagnetism for $M=441$ finite-size systems; the $M=16$ mesh overstates the van Hove singularity, see Fig. \ref{fig:Fig_kin}. In a vicinity of half-filling ferromagnetism is predicted for $M=441$ but the energy of SD state with $S=S_{\rm min}$ is not the lowest one here;
instead a state with broken translation symmetry, the three sublattice N{\'e}el state, is expected to have lower energy and competes with FM; both of these two states have been indeed observed in experiment \cite{HubbardCornell,Xu_Ferromagnetism}. For stronger interactions, ferromagnetism is predicted in a vicinity of $\nu=0.75$ for both meshes. 
In the following sections, we focus on estimates of transition temperatures 
for ferromagnetism around this particular filling, indicated by a black dashed line in Fig.~\ref{fig:SpinvsFill}.   

\subsection{Magnon energies}

\begin{figure}[h!]
\centering
\includegraphics[width=\linewidth]{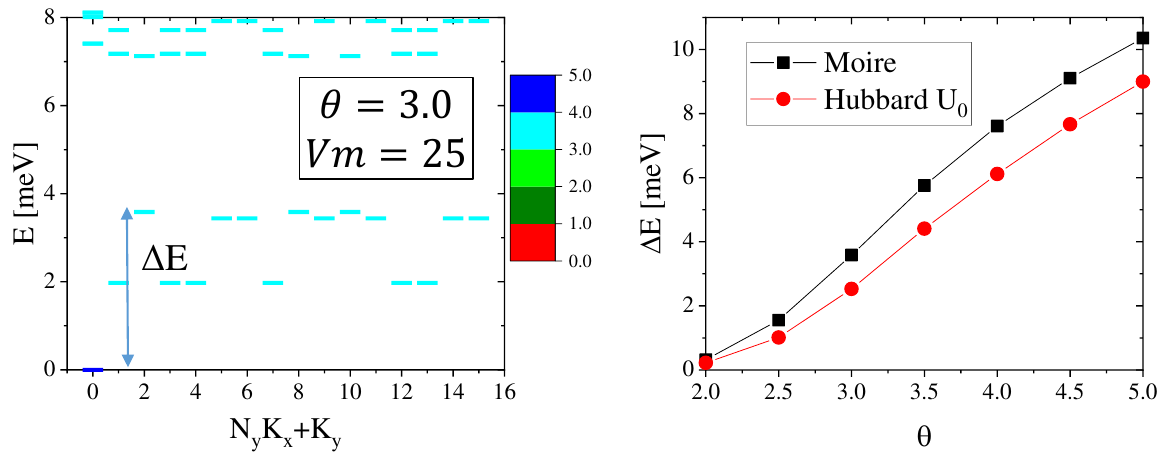}
\caption{Spin-flip excitation spectrum of a $M=N_{\rm x}\times N_{\rm y} = 16$ fully polarized ground state.
(a) Energy spectrum for total spin $S=S_{{\rm max}}-1$ for the system with $N_{{\rm h}}=9$ holes ($N=23$, $\nu=0.73$), corresponding to the filling factor $\nu$ indicated by a dashed line in Fig. \ref{fig:SpinvsFill}, $\epsilon^{-1} = 0.04$, $\psi=-94^{\circ}$. 
The 16 lowest energy excitations can be associated with magnon collective modes, and the higher energy 
excitations with unbound spin-flip particle-hole excitations. $\Delta E$ indicates the width of the magnon spectrum, which scales
with the transition temperature. $N_{\rm x}$ ($N_{\rm y}$) is the number of unit cells along two directions determined by real space lattice vectors ${\bf a}_1$ and ${\bf a}_2$ on a triangular lattice, $K_{\rm x}$($K_{\rm y}$) are total momenta along two directions determined by reciprocal space lattice vectors ${\bf b}_1$ and ${\bf b}_2$. (b) The width of magnon spectrum $\Delta E$ as a function of a twist angle $\theta$ for the moiré superlattice and its corresponding Hubbard model.}
\label{fig:magnon}
\end{figure}

In metallic ferromagnets with large splitting between majority spin and minority spin 
quasiparticle energies, the ordering temperature is typically limited by collective thermal fluctuations.
The Curie temperature then scales with the energies of the magnon modes, just as it does in 
insulating magnets. In Fig.~\ref{fig:magnon}(a) we show the 
spin-flip excitation spectrum of a typical maximally spin-polarized state near $\nu=3/4$.  We 
associate the 15 lowest energy excitations (one for each non-zero momentum) with magnon modes and 
the higher-energy excitations with unbound quasiparticle spin-flip excitations.  
We see that the magnon energies are several times smaller than the quasiparticle 
spin-splitting energy.  In Fig.~\ref{fig:magnon}(b) we show the twist angle dependence of the 
highest magnon energy, which grows with the band width, suggesting that spin-stiffness
is supplied mainly by band dispersion.

Since we neglect spin-orbit interactions, our two-dimensional 
model is spin-rotationally invariant and its critical temperature therefore vanishes (see the effect of spin-orbit interactions on a critical temperature in the SM \cite{Supplemental}).
We defer to a separate study the issue of engineering strong spin-orbit interactions in 
TMD triangular lattice moir\'e materials in order to suppress long-wavelength thermal fluctuations.
Figure~\ref{fig:magnon}(b) suggests that ferromagnetic critical temperatures approaching 
$100$ K could be achievable at large twist angles for sufficiently strong spin-orbit interactions.
However, it is important to realize that the single-band approximation could fail at large 
twist angles.  We return to this point again in the discussion section.

\subsection{Finite-temperature Lanczos method}

One of the interesting aspects of moir\'e materials physics from a fundamental point of view
is that the regime in which the temperature is comparable to or larger than the band width is experimentally 
accessible.  In the following paragraphs
we address the temperature dependence of magnetic properties over this wide energy interval.

For the  evaluation of thermodynamic properties in the canonical ensemble, we need 
to calculate thermal expectation values of relevant operators $A$: 
\begin{eqnarray}
\langle A\rangle= \frac{ \sum_{n=1}^{N_{st}}  \langle n| e^{-\beta H} A  | n \rangle }{ \sum_{n=1}^{N_{st}} \langle n| e^{-\beta H}  | n \rangle },
\end{eqnarray}
where $\beta=1/k_B T$ with $k_B$ the Boltzman constant, the 
partition function $Z =  \sum_{n=1}^{N_{st}} \langle n| e^{-\beta H}  |n \rangle $, and $| n \rangle$ is summed over
orthonormal basis states.  The exponential increase of $N_{st}$ with system size 
places severe limits on the direct application of these fundamental formulas.

The problem can be avoided if an appropriate statistical average of the full Hilbert space is generated.
In the finite temperature Lanczos method (FTLM) \cite{jaklivc1994lanczos} 
one starts with the high temperature expansion: 
\begin{eqnarray}
\langle A\rangle_{\beta \rightarrow 0}= Z^{-1} \sum_{n=1}^{N_{{\rm st}}} \sum_{k=0}^{\infty} \frac{ (-\beta)^k }{ k!} \langle n| H^k A  | n \rangle ,
\label{averA}
\end{eqnarray}
where 
\begin{eqnarray}
Z =\sum_{n=1}^{N_{{\rm st}}} \sum_{k=0}^{\infty} \frac{ (-\beta)^k }{ k!} \langle n| H^k  | n \rangle .
\label{Z}
\end{eqnarray}
The Lanczos algorithm is an iterative method for finding extreme eigenvalue of a large matrix
in which expectations of high powers of the Hamiltonian naturally appear. 
During Lanczos iteration steps,  a set of orthogonal basis vectors is generated (a Krylov space), 
spanning a finite-size space that contains approximations to 
eigenvectors corresponding to extreme eigenvalues of a full Hilbert space with accuracy 
controlled by the number of iteration steps.
In the Lanczos method the Hamiltonian is  diagonalized in this Krylov space obtaining Lanczos eigenvectors $|l\rangle$   
and the associated Lanczos energy eigenvalues $\epsilon_l$. When the number of Lanczos steps $N_l \geq k $ one can write
\begin{eqnarray}
\langle n|H^k A |n\rangle \approx  \sum_{l=1}^{N_L} \langle n| H^k| l(n)\rangle \langle l(n) | A |n\rangle = \nonumber \\ 
\sum_{l=0}^{N_L} (\epsilon_{l(n)})^k \langle n| l(n)\rangle \langle l(n)| A |n\rangle
\label{HkA} 
\end{eqnarray}
and 
\begin{eqnarray}
\langle n|H^k |n\rangle \approx  \sum_{l=1}^{N_L} (\epsilon_{l(n)})^k|\langle l(n)|n\rangle |^2  .
\label{Hk}
\end{eqnarray}
$N_L$ is a parameter of the approximation that needs to be large enough to reach accurate 
extremal energy eigenvalues;
for the calculations we present below we take $N_L = 150$. 
Inserting Eqs. \eqref{HkA} and  \eqref{Hk} into Eq. \eqref{averA} and Eq. \eqref{Z} and 
replacing the sum over all orthonormal basis states by a much smaller 
sum over $R$ random  Lanczos seed states, in analogy to Monte Carlo methods, yields
\begin{eqnarray}
\langle A\rangle \approx Z^{-1}\frac{N_{\rm st}}{N_{\rm R}}  \sum_{\nu \in N_{{\rm R}}} \sum_l^{N_{\rm L}}e^{-\beta \epsilon_{l(\nu)}} \langle l(\nu)| A|\nu\rangle  \langle \nu |l(\nu)\rangle,
\end{eqnarray}
where the partition function is
\begin{eqnarray}
Z\approx \frac{N_{st}}{N_{{\rm R}}}  \sum_\nu^{N_{{\rm R}}} \sum_l^{N_L}e^{-\beta \epsilon_{l(\nu)}} |\langle l(\nu)|\nu\rangle |^2.
\end{eqnarray}
The exponential-size Hilbert space of the Hamiltonian is thereby approximated by its spectral representation in a Krylov space spanned by the $N_{\rm L}$ Lanczos vectors starting from each random vector. The chosen random vectors $|\nu\rangle$ should ideally be mutually orthogonal, but for
practical purposes this  is not  really necessary since two vectors with random components in a large dimensional space
are always nearly orthogonal. 

\begin{figure}[h]
\centering
\includegraphics[width=\linewidth]{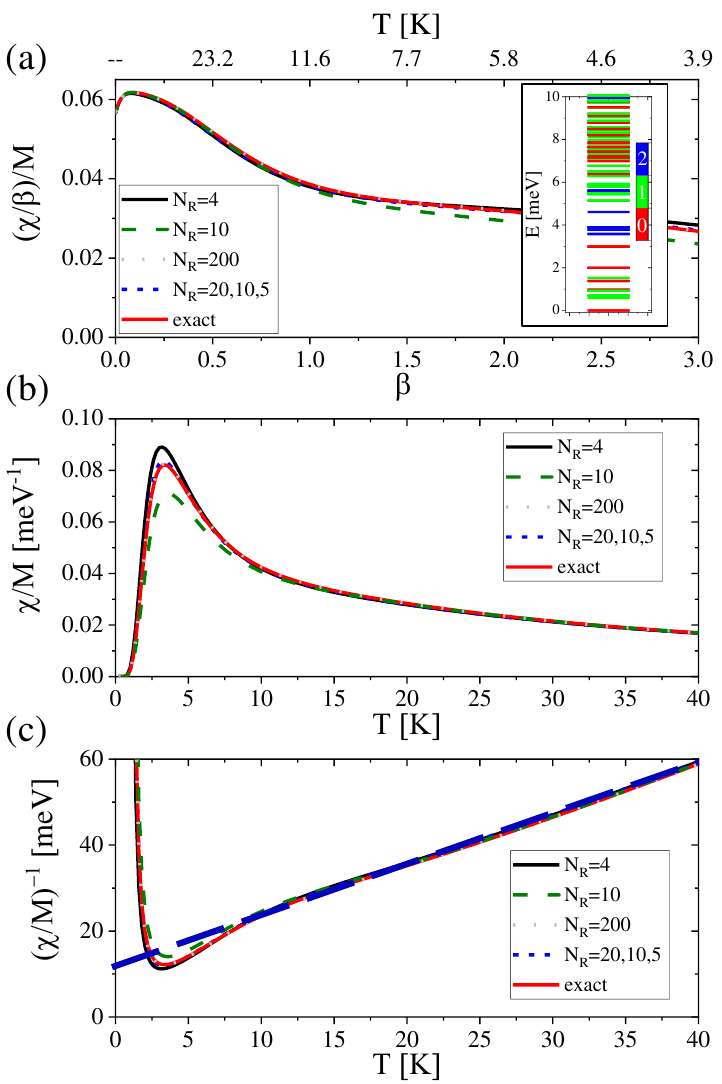}
\caption{Comparison of FTLM and exact susceptibility calculations for $M=16$ moir\'e unit cells, 
$N=4$ electrons, and different number of random vectors $N_{{\rm R}}$. (a) $\chi / \beta $ as a function of inverse temperature $\beta$. 
The inset shows low energy many-body spectrum with total spin indicated by color. (b) susceptibility $\chi (T)$ and (c) inverse susceptibility $\chi^{-1}(T)$ as a function of temperature $T$. 
The blue line in (c) is a linear fit to estimate a transition temperature $T_C$. 
The number of Lanczos steps is taken to be $N_{\rm L} = 150$. The size of the Hilbert space for $S_z = 0$ and fixed total momentum ${\bf K}$ is weakly
momentum dependent and around 900. The parameters for this illustration are 
interaction strength $\epsilon^{-1} = 0.04$, twist angle $\theta = 3.0$, $V_{{\rm m}} = 25$ meV, and $\psi=-94^{\circ}$. $N_{{\rm R}}=20,10,5$ means that for $S_{\rm z}=0$ we take  $N_{{\rm R}}=20$, $S_{\rm z}=\pm 1$ we take  $N_{{\rm R}}=10$ and so on. If one numbers is given, for all subspaces we take the same $N_{{\rm R}}$}.
\label{fig:Fig_FTLM4e}
\end{figure}

In general calculations using this approach are less sensitive to finite size effects as temperature 
increases, and most sensitive to finite size at $T=0$.
This property is related to the fact that at $T=0$ both static and dynamical quantities are calculated 
from one eigenstate only, and the selection of this state can be dependent on the size and on the shape of the 
finite-size system. 
$T > 0$ introduces thermodynamic averaging over a larger number of eigenstates and this directly reduces 
finite-size effects for static quantities.  Calculational efficiency can be 
improved by taking symmetries into account, so that $N_{st}$ corresponds to the number of states with a given symmetry.

In our view, the finite temperature Lanczos method (FTLM) is ideally suited to exploring the 
high-temperature physics that is observable in moir\'e materials.
In this work we focus on calculations of the spin magnetic susceptibility $\chi = \beta \langle S_z^2\rangle$ where 
\begin{equation}
    \langle S_z^2\rangle = \frac{\sum_n \exp{(-\beta \epsilon_n)} S_z(n)^2}{\sum_n \exp{(-\beta \epsilon_n)}}.
\end{equation}
Beacuse $[H,S_z] = 0$, the Lanczos method can be applied to each $S_z$ sector separately.  
The FTLM formula for the susceptibilty is 
\begin{eqnarray}
\chi = Z^{-1}\beta \sum_s  \frac{N_{\rm st}(s)}{N_{{\rm R}}(s)}\sum_{\nu=1}^{N_{{\rm R}}(s)} \sum_{l=1}^{N_L}e^{-\beta \epsilon_{l(\nu)}} |\langle l(\nu)|\nu\rangle |^2 s^2 ,
\end{eqnarray}
where $s$ is the $S_z$ value for the subspace. 
We find that the most accurate results are obtained for $N_{{\rm R}}(s)$ chosen such that the ratio between the Hilbert subspace size and the number of vectors is kept constant. 

\begin{figure}[h]
\centering
\includegraphics[width=\linewidth]{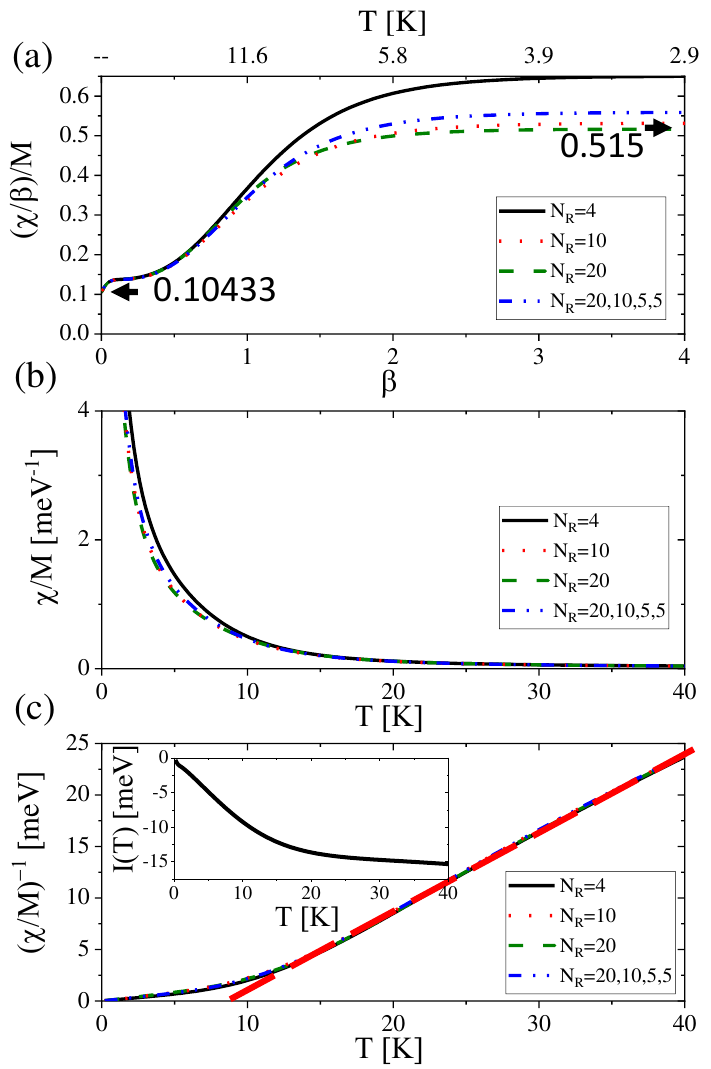}
\caption{FTLM susceptibility calculations for $M=16$ moir\'e unit cells, 
$N=23$ electrons (9 holes), and different number of random vectors $N_{{\rm R}}$.
(a) $\chi / \beta$ normalized per moir\'e cell as a function of inverse temperature $\beta$.
Analytically calculated values are given for the limit of infinite temperature (left) and zero temperature (right). 
(b) Susceptibility $\chi (T)$ and (c) inverse susceptibility $\chi^{-1}(T)$ as a function of temperature $T$. The red line is a linear fit to estimate the transition temperature $T_C$. The inset shows the Stoner parameter $I(T) \equiv \chi^{-1}(T)/M - \chi_{\rm NI}^{-1}(T)/M$, where ${\rm NI}$ means non-interacting.
The number of Lanczos steps was $N_{\rm L} = 150$. The size of Hilbert space for $S_z = 0$ at fixed total momentum ${\bf K}$ is around $5\times 10^5$. $N_{{\rm R}}=20,10,5,5$ means that for $S_{\rm z}=\pm 0.5$ we take  $N_{{\rm R}}=20$, $S_{\rm z}=\pm 1.5$ we take  $N_{\rm R}=10$ and for all other subspaces $N_{{\rm R}}=5$. If one numbers is given, for all subspaces we take the same $N_{{\rm R}}$. The parameters used for this calculation 
are: dielectric constant $\epsilon^{-1} = 0.04$, twist angle $\theta = 3.0$, $V_{{\rm m}} = 25$ meV, 
and $\psi=-94^{\circ}$.}
\label{fig:Fig_FTLM9h}
\end{figure}

The accuracy of FTLM finite-size calculations is assessed in Fig. \ref{fig:Fig_FTLM4e} 
by comparing $\chi$ as calculated by performing the full sum over all states with the FTLM sum.   
The three plots in Fig. \ref{fig:Fig_FTLM4e} ((a) $\chi / \beta$ as a function of inverse temperature $\beta$, (b) the susceptibility $\chi (T)$ and the 
(c) inverse susceptibility $\chi^{-1}(T)$ as a function of temperature $T$) emphasize different aspects of the 
temperature dependence of $\chi$.
The $\beta \to 0$ ($T \to \infty$) and $\beta \to \infty$ ($T \to 0$)
limits of $\chi/\beta$ can be calculated analytically 
by averaging $S_z^2$ over the full Hilbert space ($\chi^{-1}/\beta \to M\nu(1-\nu)/2$ for $\beta \to 0$) and over the ground state spin 
multiplet ($\chi^{-1}/\beta \to S(S+1)/3$  for $\beta \to \infty$) respectively.  
For the test case ($M=16$ and $N=4$) illustrated in Fig. \ref{fig:Fig_FTLM4e}, the 
susceptibility can be calculated exactly from the full many-body spectrum because the 
Hilbert space dimension for a given $S_z$ subspace does not exceed 1000. The exact result is indicated by a red line in 
Fig. \ref{fig:Fig_FTLM4e}, and compared with FTLM estimates based on different numbers of random vectors $N_{{\rm R}}$. 
All lines overlap for temperatures $T > 20$ K, demonstrating the high accuracy of the method in the high temperature limit. 
The ground state of the system in this case has $S=0$ (see the inset), which leads to vanishing susceptibility in (b) and divergence of the inverse susceptibility in (c) as $T \to 0$.  The susceptibility reaches a maximum at around $T=4$ K. 
The blue line in (c) is a high-temperature linear fit that extrapolates to a finite 
value for $T=0$, consistent with a paramagnetic state. 

The FTLM estimates have the advantage that they 
can be drawn from larger Hilbert spaces.  In Fig. \ref{fig:Fig_FTLM9h} we show a typical result obtained for $N=23$ particles ($N_{\rm h}=9$ holes) in the $M=16$ case, in the regime of filling factors where ferromagnetism is 
expected on the basis of the many-body ground state calculations.  In this case the many-body ground state has non-zero total spin $S=9/2$. The exact value of $\chi / M \beta$ normalized per moir\'e unit cell in the $\beta \rightarrow\infty$ limit is therefore $0.515$, as indicated by a black arrow in Fig. \ref{fig:Fig_FTLM9h}(a).
(The $\beta \to 0$ limit $0.10433$, which is independent of interactions, is also indicated by a black arrow.) 
We see that the FTLM method gives accurate results in both limits, irrespective of $N_{{\rm R}}$ at $T\rightarrow\infty$ 
and for $N_{{\rm R}} \geq 10$ at $T=0$. Generally speaking, the ratio between $N_{{\rm R}}$ and the dimension of a given Hilbert subspace is a good accuracy indicator.  The increase in $\chi$ at intermediate temperatures relative to the 
high-temperature limit shows that on average interactions lower the energies of states with larger 
$S_z$ relative to those with smaller $S_z$. The linear fit to the 
inverse susceptibility shown in Fig. \ref{fig:Fig_FTLM9h}(c) estimates the Curie temperature $T_{\rm C} \approx 9$ K
for this case, and the estimate is not strongly affected by $N_{{\rm R}}$ in reasonable ranges. The inset shows the finite size Stoner parameter $I$ that has the expected linear-in-$T$ dependence up to around $T\approx 12$ K.

Having established the efficacy of the FTLM, we now employ it to study trends in ferromagnetism in 
triangular lattice moir\'e materials.  In the SM \cite{Supplemental} in  Figs. S2(b) and S2(d) we compare inverse susceptibility results for two other 
twist angles for the same moir\'e modulation potential.  Extrapolating from high temperatures where 
finite-size effects are less severe, we see that the susceptibility at higher temperatures decreases with twist angle.
We attribute this decrease to an increase in bandwidth, which decreases the Pauli susceptibility of non-interacting 
electrons.  At the same time, the high-temperature estimate of the Curie temperature at which the susceptibility diverges
(the inverse susceptibility vanishes) increases with twist angle.  We attribute this increase also to increasing bandwidth,
which increases magnon energies by increasing the kinetic energy cost of spatial modulation of the magnetization.

\section{Discussion}
\label{sec:discussion}
We have used three different indicators available from finite-size exact diagonalization calculations to 
address the physics of itinerant ferromagnetism in single-band triangular-lattice moir\'e materials: (i)
ground state spin quantum numbers, (ii) magnon excitation energies, and (iii) temperature dependent 
spin-susceptibilities.  All indicate that ferromagnetism is common at hole band filling factors 
near $\nu=3/4$ at temperatures up to $\sim 10$K.    
Our calculations were performed for particular values of the moir\'e modulation strength and shape parameters.  
These are however expected to be strongly dependent on the specific heterojunction at which the moir\'e pattern is formed, 
and in particular on strain relaxations at those heterojunctions which will tend to increase modulation 
strengths \cite{DeepMoire_Columbia,wang2023fractional}.  When $V_m \to \lambda V_m$, twist angle $\theta \to \sqrt{\lambda} \theta$, and 
dielectric screening parameter $\epsilon \to \sqrt{\lambda} \epsilon$, the three terms in the continuum model 
Hamiltonian (interaction, moir\'e potential, and kinetic energy) all increase by a factor of $\lambda$.
Since the properties of interest here are relatively insensitive to the interaction strength parameter 
within reasonable ranges, it follows that the properties of systems with stronger moir\'e potentials 
can be read off from our results by increasing temperature scales and twist angles.
In particular, the larger energy scales increase the temperatures at which ferromagnetism can occur. 

It is interesting to compare TMD triangular lattice moir\'e materials, with graphene multilayer moir\'e materials that also support ferromagnetic states.
In the latter case, it is known that because of topological obstructions inherited from the individual layer Dirac cones \cite{Topo_Obstruction1,Topo_Obstruction2,Topo_Obstruction3}, a faithful representation of 
the flat moir\'e minibands requires multi-band tight-binding \cite{HeavyFermions_Bernevig} models, for which the exact diagonalization approach is not practical.  In the TMD moir\'e material case, however, the lowest energy 
moir\'e bands have Wannier functions that are similar to harmonic oscillator ground states centered 
on moir\'e potential extrema \cite{wu2018hubbard}.  Although we do not approximate the interaction matrix elements in our 
one-band model, we have verified that all properties related to ferromagnetism are similar to those of 
simple triangular lattice Hubbard models. 

It is also interesting to compare TMD triangular-lattice materials with rhombohedral graphene 
multilayers \cite{Young_BilayerGraphene,Young_rhombTrilayer, Young_TrilaySupercond, PhysRevB.107.L121405, Lee2022, Weitz_BilayerGraphene, han2023correlated, PhysRevB.108.L041101, das2023unconventional}, a class of two-dimensional materials in which metallic ferromagnetism has been discovered recently.  These graphene multilayer systems are like
TMD moir\'e materials in that they have peaks in their densities of states, related in that
case to Liftshitz transitions of distorted Dirac cones, but they do not have minibands and are 
not approximated by Hubbard models. The magnetism that appears in these systems is consistent with the 
notion that the key to ferromagnetism is a sharp density-of-states peak in a low-density-of-states 
background.    

At the mean-field level, the critical temperature of the 
ferromagnetic state is proportional to the exchange splitting $\Delta_{\rm exch}$
between majority and minority spin bands.  The classic metallic ferromagnets, like cobalt, iron or nickel, are well known to have transition temperatures $T_{\rm c}$ that are much lower than 
the exchange splitting $\Delta_{\rm exch}$.  Measured critical 
temperatures are more comparable to typical magnon energies $E_{\rm mag}$ 
($k_B T_{\rm c} \sim E_{\rm mag} \ll \Delta_{\rm exch} $).  Critical temperature estimates 
based on fermionic mean-field approximation do not work well for itinerant ferromagnets, actually in agreement with our results. We believe that our Hubbard model systems are, in this
sense, in the same regime as the classical $3d$ ferromagnets.

The exact diagonalization method we have employed is most suitable when the many-electron Hilbert space 
can be truncated to a single moir\'e miniband.  The small parameter which controls the applicability of this 
approximation is the ratio of the largest interaction scale, the on-site Hubbard interaction $U_0$,
to the sub-band separation.  As explained in Ref.\;\onlinecite{wu2018hubbard} these can be estimated by 
making a harmonic approximation for the moir\'e potential.  We find that 
$U_0 \sim \rm{Ry}^{3/4} (zV_m)^{1/4} (a_{B}/a_M)^{1/2}$, where $z=6$ is the triangular 
lattice coordination number and $\rm{Ry}$ and $a_{B}$ are the 
host 2D semiconductor Rydberg energy scale $\sim 0.3$ eV and Bohr radius length scale $\sim 1$ nm. 
Similarly the subband separation $\hbar \omega \sim \rm{Ry}^{1/2} (zV_m)^{1/2} a_B/a_M$.
It follows that 
\begin{equation}
    \frac{U_0}{\hbar \omega}  \sim (\rm{Ry}/zV_m)^{1/4} (a_M/a_B)^{1/2}.
    \label{eq:chargetransfer}
    \end{equation}
Truncation to the lowest moir\'e band is justified at all band filling factors 
$\nu \in (0,1)$ when the right hand side of Eq.~\ref{eq:chargetransfer} is 
smaller than $\sim 1$.  Most systems \cite{LiangFuQuantumChemistry, LiangFuMonteCarlo} that have been studied to date do not satisfy this criterion.
Since continuum model approximations are valid only for $a_M \gtrsim a_B$,
it follows that single-band ferromagnetism will occur only when
the first factor on the right side of Eq.~\ref{eq:chargetransfer} is made small,
for example by increasing the dielectric screening environment of the moir\'e system to decrease 
${\rm Ry}$, or by choosing a system with a particularly large value of $V_m$. From exponentially localized Wannier functions obtained for the topmost valence band used in our calculations, we get, for $\theta=3.0$, $U_{\rm 0}\epsilon \approx 1121$ meV \cite{Nonlocal}, $\hbar \omega \approx 58.5 $ meV. For $\epsilon^{-1} = 0.1$, $\frac{U_0(\epsilon^{-1} = 0.1)}{\hbar \omega} > 1$, while for $\frac{U_0(\epsilon^{-1} = 0.04)}{\hbar \omega} < 1$.  Thus for the limit of weaker interaction strength the single band approximation is justified. This suggests that our predictions are relevant for 
systems with sufficiently close nearby gates.

We note that Coulomb repulsion will increase the energy of the lowest energy hole miniband, as it is filled, by more than it 
increases the energies of states in higher energy moir\'e minibands.  For this reason the regime of parameter 
space in which occupation of higher energy minibands can be neglected decreases as band filling factor increases.
When correlations are included, the ground state at hole filling factor $\nu=1/2$ is often an insulator.
When its lowest energy hole-charged excitation is dominantly in a higher hole miniband, the insulator is referred to as a 
charge transfer insulator \cite{LiangFuQuantumChemistry,LiangFuMonteCarlo}. 
Since single-band ferromagnetism is most likely near 
band-filling factor $\nu=3/4$, the present single-band study is never relevant when the ground state of the half-filled 
band is a charge transfer insulator, which already involves higher energy subbands in an essential way.
If systems could be realized in which the sign of $V_m$ is reversed
(or equivalently $\psi \to \psi + 180^{\circ}$), ferromagnetism would be expected for minibands that 
are less than half-filled.  For the standard sign of $V_m$ however,
any ferromagnetism that occurs when the interaction parameter that is the subject of 
Eq.~(\ref{eq:chargetransfer}) is large, must be of multi-band character. We leave the analysis of this situation for a future study, for it requires a different approach.  

The authors acknowledge helpful interactions with L. Fu and Y. Zhang.
This work was supported by the U.S. Department of Energy, Office of Science, 
Basic Energy Sciences, under Award $\#$ DE-SC0022106. PP acknowledges support from the
Polish National Science Centre based on Decision No. 2021/41/B/ST3/03322. 
We acknowledge the Texas Advanced Computing Center (TACC) at The University of Texas at Austin for providing high-performance computer resources.

\bibliography{ItinFerro}
\onecolumngrid

\section*{Supplemental materials}


\section{Exact diagonalization Method}
The model we study has orbital and spin degrees of freedom, discrete triangular lattice translational symmetry, $SU(2)$ spin-rotational invariance, and no spin-orbit coupling (see analysis of spin-orbit coupling term on a tranistion temperature $T_{\rm c}$ in Section V).
The Hilbert space can be divided into smaller subspaces with total momentum ${\bf K}$ with discrete values determined by the number of Moire unit cells $M$, total spin ${\bf S}$, and azimuthal spin $S^z$. The basis is constructed in an occupation number representation, distributing particles among single-particle states
labeled by spin $z$ azimuthal quantum number and quasi-particle $(k_x,k_y)$ momentum. 
The total number of possible configurations $N_{\rm st}$ for particles distributed on $M$ single particle states with a given spin $N_\downarrow$ or $N_\uparrow$ is determined by a product of binomial coefficients,  $N_{\rm st}=\binom{M}{N_{\uparrow}} \cdot \binom{M}{N_{\uparrow}} $. 
The many-body Hamiltonian projected to a given total momentum ${\bf K}$ is diagonalized in 
$S^z$ subspaces. We do not rotate the Hamiltonian matrix to a ${\bf S}$ basis as this is an additional computational cost, and instead determine the ground state from calculations of expectation value of total total spin $S$ for each energy eigenstate.

A high symmetry of a triangular lattice induces a six fold degeneracy of single particle states. This affects finite-size ED results leading to spurious results at particular fillings; we refer to spin polarization around $\nu=0.25$ for $M=16$ as one of these effects. A typical finite-size sample is obtained by taking a given number of Moire unit cells along two real space primitive unit vectors and the corresponding plaquette in real space is spanned by the vectors $N_1\,{\bf a_1}$ and $N_1\,{\bf a_1}$, where ${\bf a_i}$ are the real space lattice vectors. The points forming the momentum space mesh are given by ${\bf k}=n_1\, {\bf b_1}/N_1+n_2\, {\bf b_2}/N_2$, where ${\bf b_i}$ are the reciprocal lattice vectors and $n_i$ are integers between $0$ and $N_i-1$. Such meshes sustain the symmetry of the lattice  

\begin{figure}[h]
\centering
\includegraphics[width=0.6\linewidth]{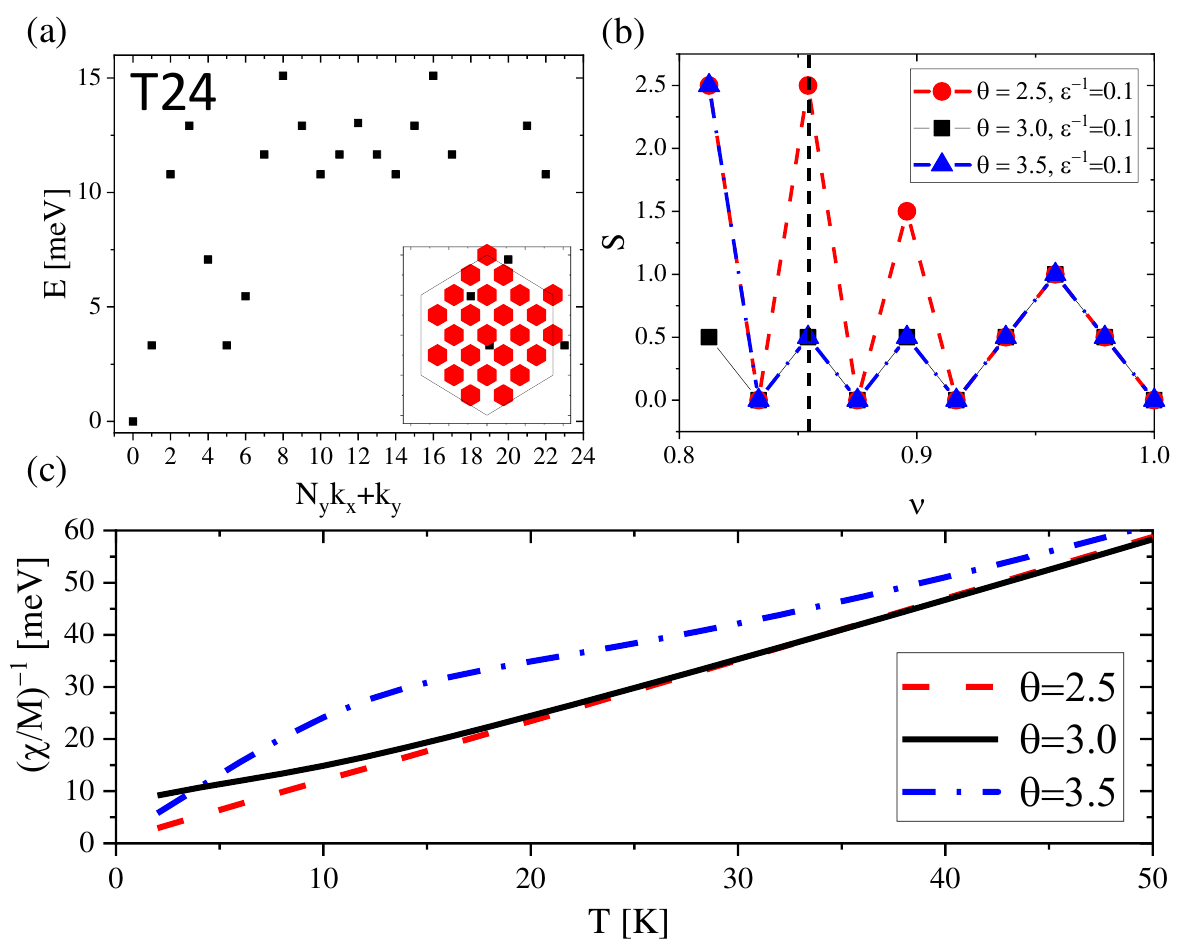}
\caption{Tilted momentum space mesh $T24$, ED and FTLM results. (a) Energies of finite size momentum space meshes for $M=24$ for $\theta = 3.0$. Insets show momentum space grid in the first Brillouin zone. (b) Total spin as a function of filling factor $\nu$ for twist angles $\theta = 2.5$, $\theta = 3.0$, and $\theta = 3.5$. (c) Inverse susceptibility $\chi^{-1}(T)$ as a function of temperature for the system with $N=7$ holes indicated by a dashed line in (b). Parameters for moiré superlattice are: dielectric constant $\epsilon^{-1} = 0.1$, $V_{{\rm m}} = 25$ meV. 
 Parameters for FTLM: $N_{\rm L} = 150$, $N_{{\rm R}} = 4$.}
\label{fig:Fig_Tilted}
\end{figure}

An alternative way of getting finite-size plaquettes is using tilted meshes \cite{TiltedM, AvellaSpringer} (see Appendix A in Ref. \onlinecite{PhysRevB.90.245401}  for a nice introduction to derivation of the tilted periodic
boundary conditions constrains). In this case, momentum space points from the interior of the Brillouin zone do not have to occur in multiplies of six.  An example of a tilted momentum space mesh with $M=24$, we call it T24, is shown in the inset in Fig. \ref{fig:Fig_Tilted}(a) with single particle energies in Fig. \ref{fig:Fig_Tilted}(a). Single particle states are non-degenerate, double, or four fold degenerate. Van Hove singularity is represented by a set of states in an energy range between $E=10$ meV and $E=13$ meV. We populate these states by up to $N_h=9$ holes; we are still limited by a maximum size of a dimension of the Hilbert space subspace. The corresponding total spin ground states for three different twist angles are shown in Fig. \ref{fig:Fig_Tilted}(b). For fillings $\nu<0.9$, finite spin polarization occurs for the smallest considered twist angle $\theta = 2.5$ for the system with odd number of particles, $N_h=5$, $N_h=7$, and $N_h=9$. This is again a finite size effect (similar finite size effects were recently noticed in finite temperature Lanczos method in Ref. \cite{FTLM_Vafek}), as one can see that below the two highest energetically single particles states, there are five closely lying states, one single and four degenerate states. In order to minimize the total energy, one can populate these two states by two or four holes, and singly occupied five states by particles with the same spin in order to maximize exchange interaction. For $N_h=7$ and $N_h=9$, the ground state total spin is $S=5/2$, which corresponds to five particles with the same spin on these five closely lying degenerate states. Above these states, there are two sets of closely lying four degenerate states which allows scatterings from/to these states. Thus, here the correlation effects are taken into account, while they were significantly reduced in the case of $M=16$ and $N_h = 9$, where the set of nine states were separated from other single particles states by a large energy gap. While for T24, correlation effects are more realistic than for symmetric meshes, at the same time one can suspect that due to lack of the full symmetry of the lattice system, exchange interactions are reduced. Thus, we consider T24 results as the other extreme limit, in opposite to $N_1=N_2$ symmetric meshes, where factors responsible for magnetization are too strongly reduced. Indeed, finite spin polarization for the system with $N_h=7$ is quite unstable; for larger twist angles it  disappears. Notice that we consider here a limit of stronger interactions, $\epsilon^{-1}=0.1$. We confirm high sensitivity of the magnetization at this filling by looking at inverse susceptibility as a function of temperature, what is shown in Fig. \ref{fig:Fig_Tilted}(c). Inverse susceptibility at high temperature limit extrapolates to $T\approx 0$ for $\theta = 2.5$ and $\theta = 3.0$, and to negative values for $\theta = 3.5$; the magnetization at finite temperature is not predicted from these calculations.  

\section{Twist angle dependence}
\begin{figure}[h!]
\centering
\includegraphics[width=0.6\linewidth]{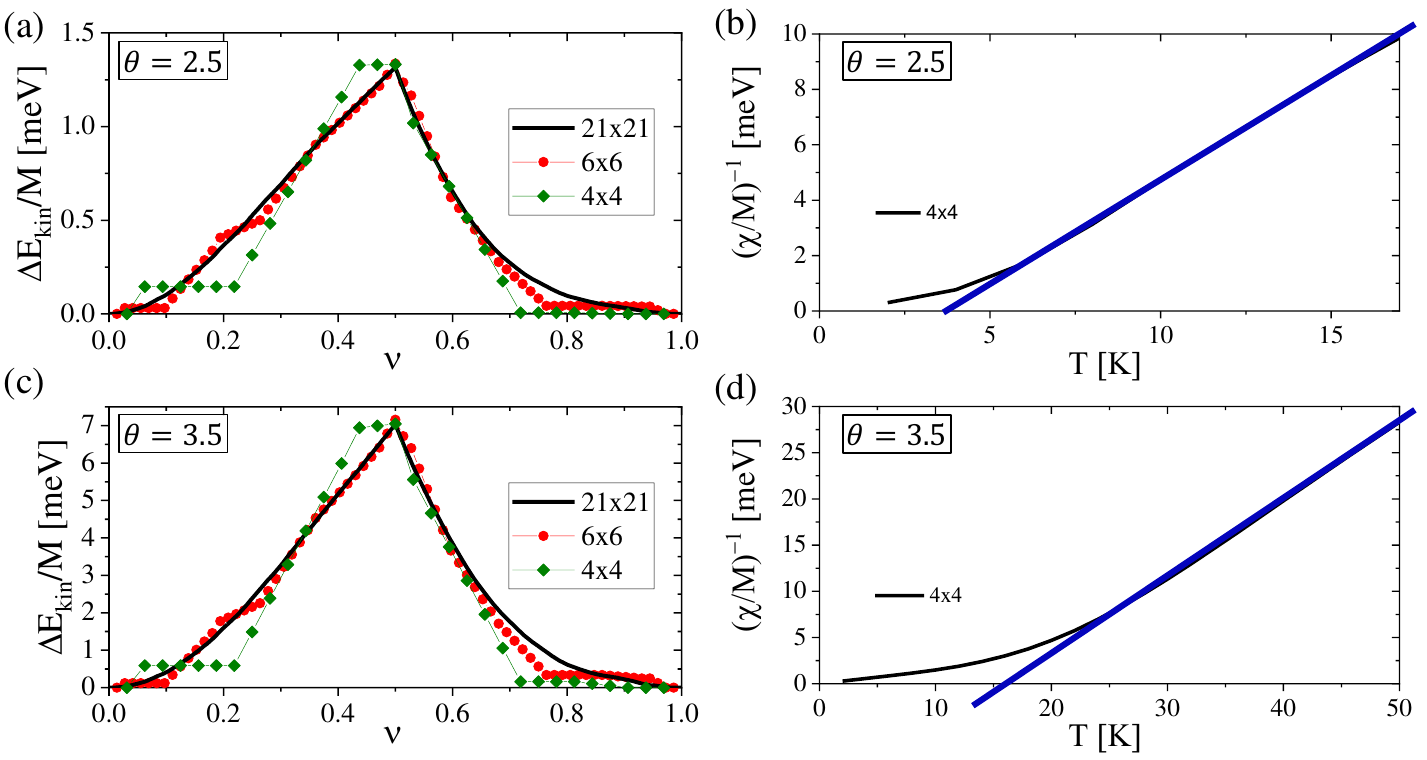}
\caption{Kinetic energy difference between maximal and minimal total spin states per moir\'e unit cell
$(\Delta E_{\rm kin}=E^{kin}_{\rm SD}(S_{\rm max}) - E^{kin}_{\rm SD}(S_{\rm min}))/M$
as a function of band filling factor $\nu=N/2M$ for different system sizes and for (a) $\theta = 2.5$ and (c) $\theta = 3.5$. Inverse susceptibility $\chi^{-1}(T)$ as a function of temperature for (b) $\theta = 2.5$ and (d) $\theta = 3.5$. Blue lines are linear fits to estimate a transition temperature $T_C$ giving $T_{\rm C} \approx 4$ K for $\theta = 2.5$ and  $T_{\rm C} \approx 16$ K for $\theta = 3.5$. The moiré superlattice parameters used for this 
calculation: interaction strength $\epsilon^{-1} = 0.04$, modulation strength $V_{{\rm m}} = 25$ meV and 
shape $\psi=-94^{\circ}$. Parameters for FTLM: $N_{\rm L} = 150$, $N_{{\rm R}} = 20, 10, 5$. }
\label{fig:Fig_kin2535}
\end{figure}

\begin{figure}[ht]
\centering
\includegraphics[width=0.6\linewidth]{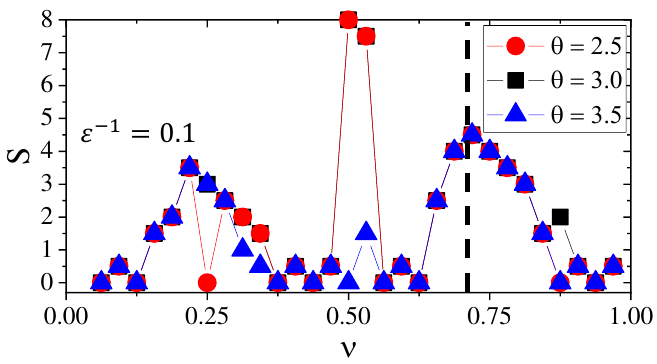}
\caption{Comparison of the ground state total spin $S$ of moiré superlattice for different twist angles for dielectric constant $\epsilon^{-1} = 0.1$ and $V_{{\rm m}} = 25$ meV, $\psi=-94^{\circ}$. Calculations for systems with $M=N_{\rm 1}\times N_{\rm 2} = 16$ unit cells. A dashed line indicates the position of the van Hove singularity for finite size mesh.}
\label{fig:SpinAng}
\end{figure}

To complement our results for twist angle $\theta = 3.0$, we analyze twist angle dependence of conditions potentially leading to ferromagnetism. We investigate the properties of the systems for $\theta = 2.5$  and $\theta = 3.5$. Figs. \ref{fig:Fig_kin2535}(a) and \ref{fig:Fig_kin2535}(c) show kinetic energy difference between maximal and minimal spin polarized states per moir\'e unit cell. For smaller (larger) twist angle, the energetic cost for spin polarization is reduced  (enlarged) because of decrease (increase) of the bandwidth. Similarity between shape of lines of energetic cost of spin polarization as a function of the filling factor in these two figures, and the one in the main article, Fig. 2, suggests that the shape of the band does not change significantly when a twist angle is increased; it is mainly nearest neighbor hopping $t$ that increases.
Changes in the bandwidth, translates into changes in transition temperatures read from $\chi^{-1}$, shown in Fig. \ref{fig:Fig_kin2535}(b) and (d). However, one needs to be caution with the band-mixing that might be significant for larger twist angles, see the Discussion section in the main article. In Fig. \ref{fig:SpinAng} we show total spin as a function of the filling factor for all three considered twist angles. In a vicinity of van Hove singularity, indicated by a dashed line, finite spin polarization  is expected for all three systems. Increase of the twist angle to $\theta > 3.0$ leads to a transition between maximially spin polarized state at half-filling, to minimal spin polarization; antiferromagnetic superexchange dominates when the bandwidth increases (hopping integral increases) over direct exchange interaction. 

\section{Hartree-Fock}
\begin{figure}[h!]
\centering
\includegraphics[width=0.6\linewidth]{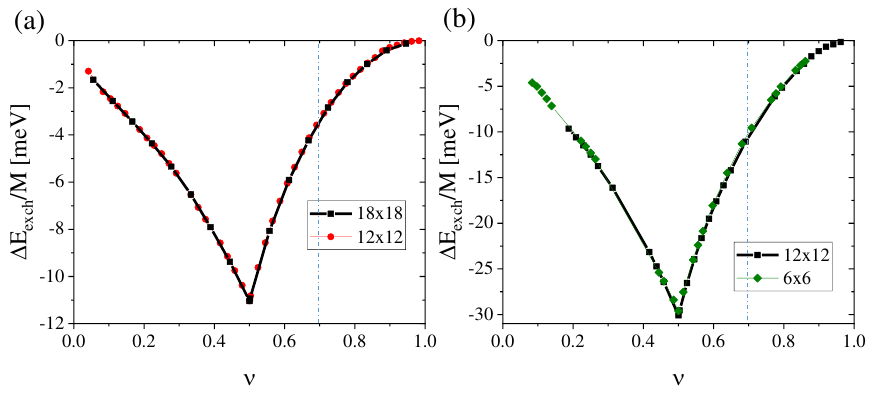}
\caption{Exchange energy difference between maximally $S_{\rm max}$ and minimally $S_{\rm min}$ spin polarized states normalized per moir\'e unit cell from Hartree-Fock self-consistent calculations for  
(a) $\epsilon^{-1}=0.04$ and (b) $\epsilon^{-1}=0.1$ interaction strength. 
The results for twist angle $\theta = 3.0$, moir\'e modulation strength $V_{{\rm m}} = 25$ meV, and potential shape $\psi=-94^{\circ}$. }
\label{fig:exchHF}
\end{figure}

We employ the self-consistent Hartree-Fock approximation to compute the exchange energy for finite-size systems exactly. Hartree-Fock theory approximates interaction through an additional self-energy term $\Sigma^{HF}$, which can be expressed using a self-consistently determined single-particle density matrix $\rho = \sum_{n} \ket{\psi_n}\bra{\psi_n}$. The sum here is over the occupied moir\'e Bloch states.
To evaluate the mean-field electronic structure of moir\'e superlattices, a plane-wave representation is employed, where the Hartree-Fock self-energy $\Sigma^{HF}$ at each 
$\b{k}$ in the Brillouin-zone is a matrix in moir\'e reciprocal lattice vectors $\b{G}$:
\begin{align}
    \Sigma^{HF}_{\alpha,\b{G};\beta,\b{G}'}(\b{k}) =& \Sigma^{H}_{\alpha,\b{G};\beta,\b{G}'}(\b{k}) + \Sigma^{F}_{\alpha,\b{G};\beta,\b{G}'}(\b{k}) \nonumber \\
    =& \frac{\delta_{\alpha,\beta}}{A}\sum_{\alpha'}V(\b{G}'-\b{G})\sum_{\b{k}',\b{G}''}\rho_{\alpha',\b{G}+\b{G}'';\alpha',\b{G}'+\b{G}''}(\b{k}')  -\frac{1}{A}\sum_{\b{G}'',\b{k}'}V(\b{G}''+\b{k}'-\b{k})\rho_{\alpha,\b{G}+\b{G}'';\beta,\b{G}'+\b{G}''}(\b{k}').
\label{eq:self-energy}
\end{align}
In Eq.~\ref{eq:self-energy}, Greek letters label spin, and $\rho_{\alpha,\b{G};\beta,\b{G}'}$ is the momentum-space density matrix with a chosen symmetry. We do not project the interaction to the lowest energy band in Hartree-Fock, all the bands are taken into account, and the Coulomb interaction takes the standard form: $V(q) = 2\pi e^2/\epsilon q$. Starting with a physically plausible density matrix $\rho_0$, we minimize the energy by performing self-consistent iterations. We consider a state to be converged when the total Hartree-Fock energy difference between iterations is less than $10^{-3}$ meV. The exchange energy is evaluated after convergence by $E_{\text{exch}}^{HF} = \Tr(\Sigma^F(\b k)\rho(\b k))/2$. Exchange energy difference between maximally $S_{\rm max}$ and minimally $S_{\rm min}$ spin polarized states normalized per moir\'e unit cell from Hartree-Fock self-consistent calculations  are shown in Fig. \ref{fig:exchHF}(a) for $\epsilon^{-1}=0.04$ and Fig. \ref{fig:exchHF}(b) for $\epsilon^{-1}=0.1$ interaction strength. Clearly, mixing with other bands enhance this difference in comparison to single Slater determinant calculations shown in Fig. 4(a) and 4(b) in MS. 

To compare with ED calculations that are projected to the lowest-energy band, the two relevant competing states for the present work are minimally spin-polarized ($\nu_{\uparrow} = \nu_{\downarrow} = \nu/2$) and maximally spin-polarized ferromagnetic ($\nu_{\uparrow} = \nu$ for $\nu \leq 1/2$; $\nu_{\uparrow}=1/2, \nu_{\downarrow} = \nu-1/2$ for $\nu>1/2$) states. Our calculations are carried out such that the particle number for each spin is conserved. We use the same scheme of discretization for the Brillouin zone and the same continuum model parameters as ED in Fig.~\ref{fig:exchHF}.

\section{Role of Long Range Interactions}
\begin{figure}[h!]
\centering
\includegraphics[width=0.8\linewidth]{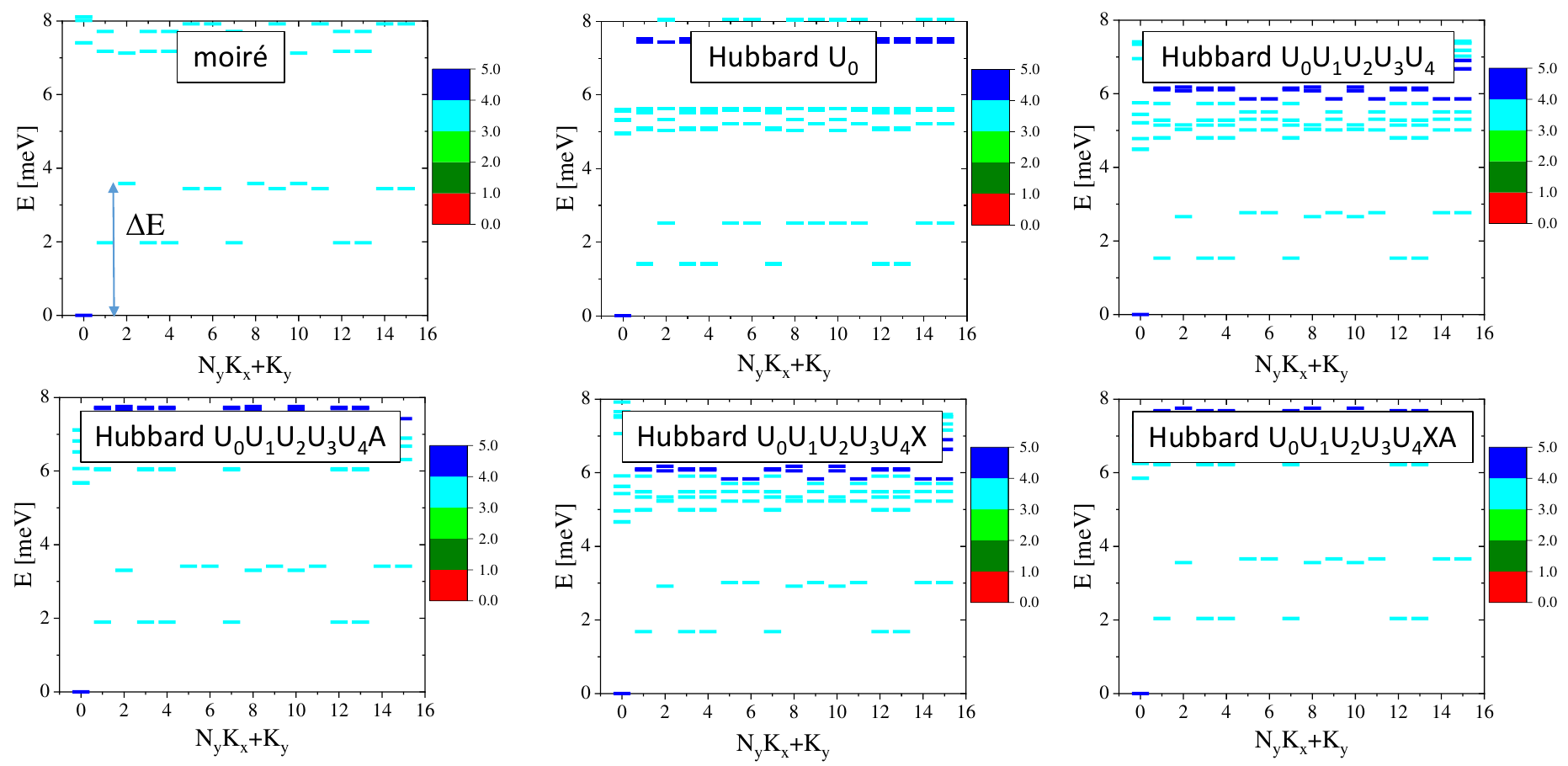}
\caption{$\theta = 3.0$ analysis. Energy spectrum for total spin $S=S_{{\rm max}}-1$ for the system with $N_{{\rm h}}=9$ holes, corresponding to filling factor $\nu$ indicated by a dashed line in Fig. 2 in MS, $\epsilon^{-1} = 0.04$, for different models. $\Delta E$ indicates the width of magnon spectrum, which determines transition temperature to a ferromagnetic phase. $N_{\rm x}$($N_{\rm y}$) is a number of unit cells along two directions determined by real space lattice vectors ${\bf a}_1$ and ${\bf a}_2$ on a triangular lattice, $K_{\rm x}$($K_{\rm y}$) are total momenta along two direction determined by reciprocal space lattice vectors ${\bf b}_1$ and ${\bf b}_2$. Parameters for moiré superlattice are: dielectric constant $\epsilon^{-1} = 0.04$, twist angle $\theta = 3.0$, $V_{{\rm m}} = 25$ meV. }
\label{fig:LongHubb30}
\end{figure}

\begin{figure}[ht]
\centering
\includegraphics[width=0.6\linewidth]{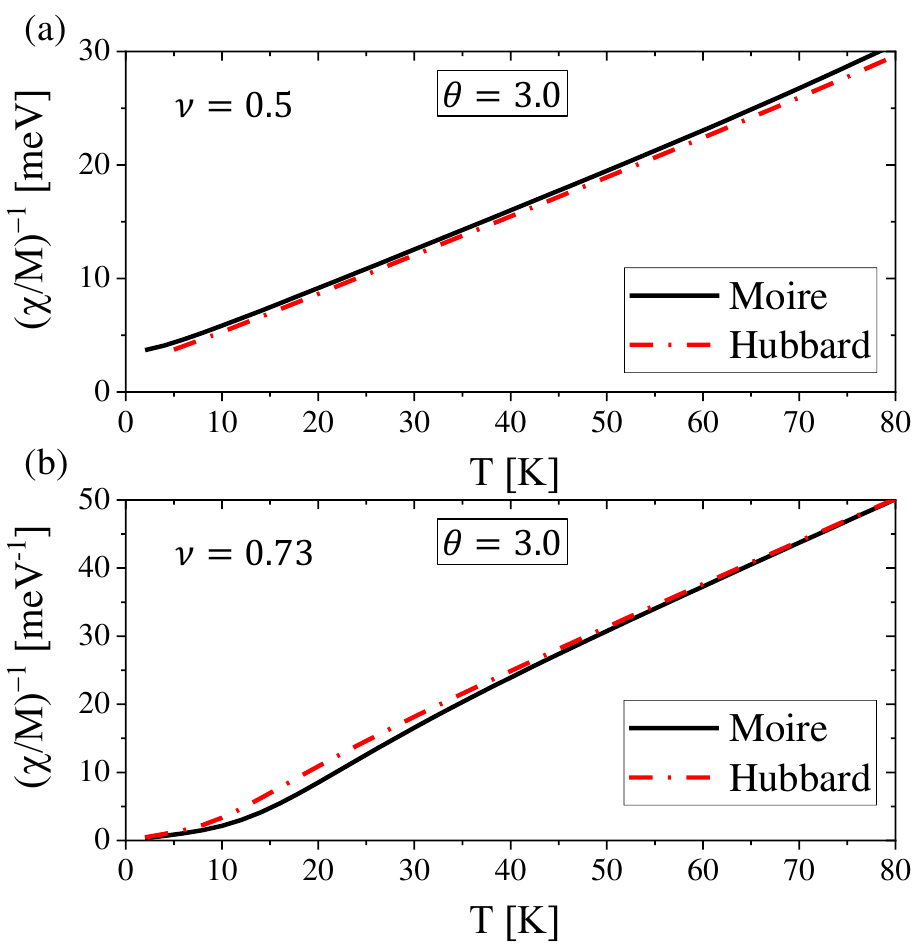}
\caption{Comparison between moiré system and Hubbard model results. Inverse susceptibility $\chi^{-1}(T)$ as a function of temperature. Calculations for $M=16$ with (a) $N=16$ electrons $\nu=0.5$ and (b) $N=23$ particles $\nu=0.73$ ($N_{{\rm h}}=9$ holes). The number of random vectors $N_{{\rm R}} = 4$. Parameters for moiré superlattice are: dielectric constant $\epsilon^{-1} = 0.04$, twist angle $\theta = 3.0$, $V_{{\rm m}} = 25$ meV.}
\label{fig:Fig_MoireHubb}
\end{figure}

Many-body model of moir\'e superlattice inherits many features from on-site Hubbard model but in general long-range and non-local interactions affects the ground state properties. This can be seen as ferromagnetism at half-filling for sufficiently strong interactions, due to direct exchange, and from formation of generalized Wigner crystal formation for other partial fillings due to long range part of Coulomb interaction. These extra terms also affect the charge distributions of spin polarized and depolarized states because the occupation of single particles states is interaction dependent. In Fig. \ref{fig:LongHubb30} we analyze how long range direct interaction, together with direct exchange and assisted hoppings, affects magnon spectrum at van Hove singularity for $M=16$ for $\epsilon^{-1} = 0.04$. The width of magnon spectrum from onsite and extended Hubbard models are $\Delta E_{\rm Hubb} = 2.8$ meV. This energy is increased by around 1 meV, when non-local interaction terms are included. This also translates to stronger tendency for ferromagnetism around van Hove singularity for moir\'e systems in comparison to onsite Hubbard model, what can be read from lower values of inverse susceptibility at low temperatures in Fig. \ref{fig:Fig_MoireHubb}(b). 

\section{Critical temperature estimation with spin-orbit coupling}
\begin{figure}[ht]
\centering
\includegraphics[width=0.6\linewidth]{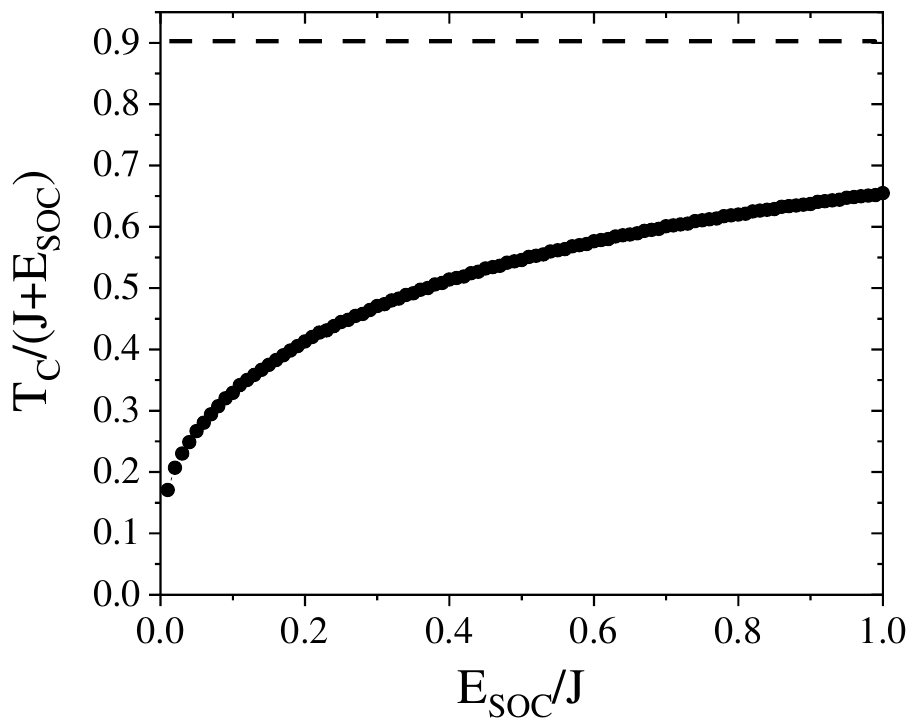}
\caption{Critical temperature $T_{\rm c}$ as a function of spin-orbit coupling strength $E_{\rm SOC}$.}
\label{fig:ESOC}
\end{figure}

We use a simplified model to estimate the effect of spin rotational symmetry breaking term on a transition temperature to a ferromagnetic phase. The magnetization drop as a function of temperature is determined by a number of excited magnons. We can write
\begin{align}
    \langle S\rangle = N/2-N_{\rm mag}(T),
\label{eq:magnet}
\end{align}
$N_{\rm mag}$ is a number of excited magnons at a given temperature $T$. The critical temperature can be estimated from a condition
\begin{align}
    N/2 = N_{\rm mag}(T_{\rm c}).
\label{eq:Tc}
\end{align}
While the exact magnon spectrum is not known, we assume a simplified model of a circular Brillouin zone (BZ) in 2D (in 3D this would be a spherical BZ) with a momentum cut-off ${\bf k}_{\rm D}$ (Debye model of density of states). In that case, a total number of particles, a normalization factor, is given by 
\begin{align}
    N/2 = \frac{A}{2\pi}\int_0^{ k_{\rm D}} kd{k},
\label{eq:Npart}
\end{align}
where $A=A_{\rm UC}M$ is are of the system, $A_{\rm UC}=a_{\rm M}^2\sqrt{3}/2$ is the area of the unit cell, $a_{\rm M}$ is moir\'e lattice vector ($A_{\rm BZ}=\frac{(2\pi)^2)}{A_{\rm UC}} = \pi k_{\rm D}^2$, what gives $k_{\rm D}^2 = 4\pi/A_{\rm UC}$). Magnon energy for a system with spin-orbit coupling is 
\begin{align}
    E_{\rm mag} = E_{\rm SOC} + D k^2,
\label{eq:Emag}
\end{align}
where $D=\frac{J}{k_{\rm D}^2}$ is spin stiffness and $J$ is magnon energy at BZ boundary, $E_{\rm SOC}$ is spin-orbit coupling strength. The total number of magnons at a given temperature is 
\begin{align}
    N_{\rm mag} = \int_0^{ k_{\rm D}} d{k} k n_{\rm B}(E_{\rm mag}(k)),
\label{eq:Nmag}
\end{align}
where $n_{\rm B}(E)=\frac{1}{\exp{(E/k_{\rm B}T)}-1}$ is the Bose-Einstein distribution. We numerically solve Eq. \ref{eq:Tc} in order to determine $T_{\rm c}$ as a function of spin-orbit coupling strength $E_{\rm SOC}$. The results are shown in Fig. \ref{fig:ESOC}. An infinitesimally small value of $E_{\rm SOC}$ is sufficient to induce a finite value of $T_{\rm c}$. A dashed line indicates an asymptotic behavior.

\end{document}